\documentstyle[12pt,amscd,amssymb]{article}

\begin{document}

\baselineskip16pt

\newtheorem{definition}{Definition $\!\!$}[section]
\newtheorem{prop}[definition]{Proposition $\!\!$}
\newtheorem{lem}[definition]{Lemma $\!\!$}
\newtheorem{corollary}[definition]{Corollary $\!\!$}
\newtheorem{theorem}[definition]{Theorem $\!\!$}
\newtheorem{example}[definition]{\it Example $\!\!$}
\newtheorem{remark}[definition]{\it Remark $\!\!$}

\newcommand{\nc}[2]{\newcommand{#1}{#2}}
\newcommand{\rnc}[2]{\renewcommand{#1}{#2}}
\nc{\bpr}{\begin{prop}}
\nc{\bth}{\begin{theorem}}
\nc{\ble}{\begin{lem}}
\nc{\bco}{\begin{corollary}}
\nc{\bre}{\begin{remark}}
\nc{\bex}{\begin{example}}
\nc{\bde}{\begin{definition}}
\nc{\ede}{\end{definition}}
\nc{\epr}{\end{prop}}
\nc{\ethe}{\end{theorem}}
\nc{\ele}{\end{lem}}
\nc{\eco}{\end{corollary}}
\nc{\ere}{\end{remark}}
\nc{\eex}{\end{example}}
\nc{\epf}{\hfill\mbox{$\Box$}}
\nc{\ot}{\otimes}
\nc{\bsb}{\begin{Sb}}
\nc{\esb}{\end{Sb}}
\nc{\ct}{\mbox{${\cal T}$}}
\nc{\ctb}{\mbox{${\cal T}\sb B$}}
\nc{\bcd}{\[\begin{CD}}
\nc{\ecd}{\end{CD}\]}
\nc{\bea}{\begin{eqnarray*}}
\nc{\eea}{\end{eqnarray*}}
\nc{\be}{\begin{enumerate}}
\nc{\ee}{\end{enumerate}}
\nc{\beq}{\begin{equation}}
\nc{\eeq}{\end{equation}}
\nc{\bi}{\begin{itemize}}
\nc{\ei}{\end{itemize}}
\nc{\kr}{\mbox{Ker}}
\nc{\te}{\!\ot\!}
\nc{\pf}{\mbox{$P\!\sb F$}}
\nc{\pn}{\mbox{$P\!\sb\nu$}}
\nc{\bmlp}{\mbox{\boldmath$\left(\right.$}}
\nc{\bmrp}{\mbox{\boldmath$\left.\right)$}}
\rnc{\phi}{\mbox{$\varphi$}}
\nc{\LAblp}{\mbox{\LARGE\boldmath$($}}
\nc{\LAbrp}{\mbox{\LARGE\boldmath$)$}}
\nc{\Lblp}{\mbox{\Large\boldmath$($}}
\nc{\Lbrp}{\mbox{\Large\boldmath$)$}}
\nc{\lblp}{\mbox{\large\boldmath$($}}
\nc{\lbrp}{\mbox{\large\boldmath$)$}}
\nc{\blp}{\mbox{\boldmath$($}}
\nc{\brp}{\mbox{\boldmath$)$}}
\nc{\LAlp}{\mbox{\LARGE $($}}
\nc{\LArp}{\mbox{\LARGE $)$}}
\nc{\Llp}{\mbox{\Large $($}}
\nc{\Lrp}{\mbox{\Large $)$}}
\nc{\llp}{\mbox{\large $($}}
\nc{\lrp}{\mbox{\large $)$}}
\nc{\lbc}{\mbox{\Large\boldmath$,$}}
\nc{\lc}{\mbox{\Large$,$}}
\nc{\Lall}{\mbox{\Large$\forall$}}
\nc{\bc}{\mbox{\boldmath$,$}}
\rnc{\epsilon}{\varepsilon}
\rnc{\ker}{\mbox{\em Ker}}
\nc{\ra}{\rightarrow}
\nc{\ci}{\circ}
\nc{\cc}{\!\ci\!}
\nc{\T}{\mbox{\sf T}}
\nc{\can}{\mbox{\em\sf T}\!\sb R}
\nc{\cnl}{$\mbox{\sf T}\!\sb R$}
\nc{\lra}{\longrightarrow}
\nc{\M}{\mbox{Map}}
\rnc{\to}{\mapsto}
\nc{\imp}{\Rightarrow}
\rnc{\iff}{\Leftrightarrow}
\nc{\bmq}{\cite{bmq}}
\nc{\ob}{\mbox{$\Omega\sp{1}\! (\! B)$}}
\nc{\op}{\mbox{$\Omega\sp{1}\! (\! P)$}}
\nc{\oa}{\mbox{$\Omega\sp{1}\! (\! A)$}}
\nc{\inc}{\mbox{$\,\subseteq\;$}}
\nc{\de}{\mbox{$\Delta$}}
\nc{\spp}{\mbox{${\cal S}{\cal P}(P)$}}
\nc{\dr}{\mbox{$\Delta_{R}$}}
\nc{\dsr}{\mbox{$\Delta_{\cal R}$}}
\nc{\m}{\mbox{m}}
\nc{\0}{\sb{(0)}}
\nc{\1}{\sb{(1)}}
\nc{\2}{\sb{(2)}}
\nc{\3}{\sb{(3)}}
\nc{\4}{\sb{(4)}}
\nc{\5}{\sb{(5)}}
\nc{\6}{\sb{(6)}}
\nc{\7}{\sb{(7)}}
\nc{\hsp}{\hspace*}
\nc{\nin}{\mbox{$n\in\{ 0\}\!\cup\!{\Bbb N}$}}
\nc{\ha}{\mbox{$\alpha$}}
\nc{\hb}{\mbox{$\beta$}}
\nc{\hg}{\mbox{$\gamma$}}
\nc{\hd}{\mbox{$\delta$}}
\nc{\he}{\mbox{$\varepsilon$}}
\nc{\hz}{\mbox{$\zeta$}}
\nc{\hs}{\mbox{$\sigma$}}
\nc{\hk}{\mbox{$\kappa$}}
\nc{\hm}{\mbox{$\mu$}}
\nc{\hn}{\mbox{$\nu$}}
\nc{\hl}{\mbox{$\lambda$}}
\nc{\hG}{\mbox{$\Gamma$}}
\nc{\hD}{\mbox{$\Delta$}}
\nc{\th}{\mbox{$\theta$}}
\nc{\Th}{\mbox{$\Theta$}}
\nc{\ho}{\mbox{$\omega$}}
\nc{\hO}{\mbox{$\Omega$}}
\nc{\hp}{\mbox{$\pi$}}
\nc{\hP}{\mbox{$\Pi$}}

\def\esl{{\mbox{$E\sb{\frak s\frak l (2,{\Bbb C})}$}}}
\def\esu{{\mbox{$E\sb{\frak s\frak u(2)}$}}}
\def\zf{{\mbox{${\Bbb Z}\sb 4$}}}
\def\zt{{\mbox{$2{\Bbb Z}\sb 2$}}}
\def\ox{{\mbox{$\Omega\sp 1\sb{\frak M}X$}}}
\def\oxh{{\mbox{$\Omega\sp 1\sb{\frak M-hor}X$}}}
\def\oxs{{\mbox{$\Omega\sp 1\sb{\frak M-shor}X$}}}

\title{\vspace*{-15mm}\sc\Large
Strong Connections on Quantum Principal Bundles
}
\author{{\sc Piotr M.\ Hajac}
\thanks{
On~leave from:
Department of Mathematical Methods in Physics, 
Warsaw University, ul.~Ho\.{z}a 74, Warsaw, \mbox{00--682~Poland}.
http://info.fuw.edu.pl/KMMF/ludzie\underline{~~}ang.html 
(E-mail: pmh@fuw.edu.pl)
}\\
Mathematics Section, International Centre for Theoretical Physics,\\
Strada Costiera 11, 34014 Trieste, Italy. 
}
\date{6 May 1996}
\maketitle

\begin{abstract}
A gauge invariant notion of a strong connection is  
presented and characterized.  
It is then used to justify the way in which a global
curvature form is defined. 
Strong connections are interpreted as those that are induced from
the base space of a quantum bundle. Examples of both strong and non-strong
connections are provided. In particular, such connections are constructed on
a quantum deformation of the Hopf fibration $S\sp2\ra RP\sp2$. A certain class of
strong $U_{q}(2)$-connections on a trivial quantum principal bundle
is shown to be equivalent to the class of connections on a free module that are 
compatible with the $q$-dependent hermitian metric. 
A particular form of the Yang--Mills action on a trivial $U\sb q(2)$-bundle
is investigated. It is proved to coincide with the Yang--Mills action constructed
by A.~Connes and \mbox{M.\ Rieffel}. Furthermore, it is shown that the moduli space
of critical points of this action functional is independent of $q$. 
\end{abstract}

\section*{Introduction}

Two of the mainstreams of Noncommutative Geometry 
concentrate around the notions of a
projective module \cite{conbook,coq} and 
of a quantum group \cite{frt,cmp}.  Quite
recently (see \cite{bmq,md1,mp}), the concept of a quantum principal 
bundle was systematically developed with quantum groups (Hopf algebras) in the
role of structure groups.
Hence, since both projective modules and quantum principal bundles 
serve as starting
points for quantum geometric considerations, 
the conceptual framework provided by the notion of a 
quantum principal bundle has a good chance of unifying 
those two branches of Noncommutative Geometry. 

In the classical differential geometry,
it is hard to overestimate the interplay between Lie groups and $K$-theory.
Therefore, it is natural to expect that establishing a similar
interaction in the noncommutative case is necessary for better understanding of
quantum geometry. 
It is already known that 
the classification of quantum principal bundles over manifolds depends only on
 the  classical subgroups of quantum 
structure groups \cite{md2}. This leads to 
the following questions: Is the classification of (general)
quantum principal bundles over
noncommutative spaces richer then the classification of classical-group bundles
over noncommutative spaces? Is there 
a bimodule that can be obtained  
as the bimodule of intertwiners (noncommutative analogue of
equivariant vector valued functions on a total space) only from a bundle with a 
noncommutative structure Hopf algebra? More generally, when does a 
deformation of a group into a quantum group entail essential consequences in the
geometry (e.g., in the classification of bundles or in the Yang--Mills theory)?

Since this article is to a great extent a
follow-up of \cite {bmq}, much of the mathematical
and physical motivations listed there can also
be considered as motivations for this work, and so
will not be repeated here. Let us just emphasize
that our main purpose is to specify and analyze a class of
connections on quantum principal
bundles (called strong connections) that enjoy
some additional properties making
them more like their classical counterparts, and
(taking advantage of the notion of a strong connection) to discuss a link
between the two approaches to noncommutative
differential geometry based on quantum principal bundles and
projective modules. The study of the precise relationship between those two
approaches is thought of as a move towards answering the questions 
mentioned above.

We begin this article by fixing the notation and recalling the fundamentals of
 quantum bundles and Yang--Mills theory on projective modules. 
In the first section, in addition
to this vocabulary review, we also study the definition
of a quantum principal bundle using snake diagrams (see Remark~\ref{snake} 
and Proposition~\ref{galois}). Taking advantage of Remark~\ref{snake}, we prove 
(Corollary~\ref{dcor}) that the fundamental vector field compatibility condition
(see Point~3 in Definition~\ref{qpbdef}) implies its stronger version. 
(The latter version of
the fundamental vector field compatibility condition was assumed in 
Example~4.11\bmq .)

The formalism 
used in this paper is a generalization of the
corresponding formalism used in classical
differential geometry. The calculations showing this, 
though often very instructive, are straightforward
 and  we will not fully elaborate on
that  fact later on. Differential geometry on
quantum principal bundles is still in the process
of being born --- the umbilical cord has hardly
been cut yet --- and it seems premature at
this point to make precise categorical
statements establishing the relationship
between  classical and quantum differential
geometries. As Yu.I.Manin mentioned in a
similar context (see p.86 in \cite{ma}), ``Here,
one should not act too hastily since even in
supergeometry this program was started only
recently and revealed both rich content and
some puzzling new phenomena."

In Section~2, 
we define and provide examples of strong connections.
Proposition~\ref{scprop}
allows one to interpret strong connections on trivial quantum bundles
as those induced from the base space, and to produce examples of strong
connections in the case of trivial quantum bundles with the universal calculus. 
In the Introduction to the preliminary version of \bmq ,
one can read regarding inducing connection forms from the
base space that ``...in the general non-commutative or quantum case there
would appear to be slightly more possibilities..." than in the classical case.
Examples of connections that are not strong (and thus realize the
 just mentioned ``quantum possibility") are supplied as well. More precisely,
we construct both strong and non-strong connections on a very simple (yet
 rich enough) example of a `discrete bundle' and on a quantum version of 
the Hopf fibration $S\sp2\ra RP\sp2$.
(As a byproduct of our considerations we obtain a \mbox{$q$-deformation}
 of~$RP\sp2$.)
It might be worthwhile to note that the latter construction does not employ
the trivial-bundle or the quantum-group-quotient techniques. We end this section
with presenting an example of a non-strong connection in the set-up of a strict
monoidal category dual to the category of sets with Cartesian product.

In the subsequent section, we describe the action of (global) 
gauge transformations on the space of connections on a bundle with
the universal differential calculus, and show that this action 
preserves the strongness of a connection.

In Section~4, we use the notion of a strong connection to justify the
definition of a global curvature form. (We need to assume that a connection 
is strong if we want to show that its curvature form has
certain properties that classical curvature
forms possess automatically; e.g., the usual relation to the square of the exterior
covariant derivative.) 

In Section~5, we present a link between gauge theory on a
quantum principal bundle and Yang--Mills theory
on a projective module. First we show how, in the case of a free module,
to incorporate the Yang--Mills action constructed in \cite{conri,ri}
into the quantum bundle picture. Then, to obtain a
Hermitian metric compatible connection
on a free module from a strong $U\sb q(2)$-connection on a
trivial quantum principal bundle, we mimic the
 classical geometry formula which permits one to
determine the values of  a connection form on a Hopf
algebra of smooth functions on a matrix Lie group
by knowing its values on the matrix of
generators. (Note that usually one thinks of a
connection form as a map sending smooth
 vector fields to elements of a Lie algebra, but
we can also view it as a map from the Hopf
algebra of smooth functions on a Lie group into
the space of smooth \mbox{1-forms}.) It turns out that the connections
compatible with a particular $q$-dependent Hermitian structure can be 
identified with the strong $U\sb q(2)$-connections that satisfy certain condition.
We close this section by concluding that, in the setting under consideration,
the moduli space of critical points of $U\sb q(2)$ and $U(2)$--Yang--Mills
theory coincide.
Thus, at least in this case, the $q$-deformation of the structure group alone
has no essential
bearing on the Yang--Mills theory. This seems to bring us a step closer 
to answering the question posed at the end of the first paragraph: 
One should expect geometrically interesting effects of the noncommutativity 
of a Hopf algebra in Yang--Mills theory
only for non-trivial bundles (comodule algebras that are not crossed product 
algebras) or non-strong connections. 

Finally, in the Appendix, we examine the
advantages of adding a twist to the definition
of a quantum associated bundle
formulated in \bmq\  and point out the
possibility of using the axiomatic definition
of a frame bundle to try to define its
noncommutative analogue (cf.\
Section~5.1 in \cite{bmq}).

\parindent0pt
\section{Preliminaries}

The notation used throughout this article is quite
standard and 
not much different from that
of \bmq . Nevertheless, to eschew any possible misunderstanding or
confusion, we enclose a table of basic notations:

\bi

\item [{$[\! ...]_{X}$}] an equivalence class
defined by $X$

\item [$\delta_{m,n}$] equals 1 iff $m=n$, and 0
otherwise (Kronecker symbol)

\item [$\frak g\;\;\;\:$] Lie algebra of a Lie group $G$

\item [$k\;\;\;\,$] field of characteristic zero 
(except for Proposition~\ref{spera})

\item [$\ot\;\;\;$] tensor product over $k$

\item [$\tau\;\;\;\,$] flip ($\tau (u\ot v):= v\ot u$)

\item [$\Omega\sp{1}\! A\;$]  first order universal
differential calculus ($\Omega\sp{1}\! A:=\kr\,
m_{A}$ , $da:=1\ot a-a\ot 1 $)

\item [$\Omega A\;\,$] differential envelope of $A$ 

\item [$\Omega (\! A)$] differential algebra over $A$ (i.e.~a quotient of $\hO A$
by some differential ideal)

\item [$\; m_{X}\;$] multiplication on $X$, or in $X$
(We will simply write $m$ for the multiplication in a Hopf algebra.)

\item [$\Delta\;\;\;$] comultiplication: $\de (a)=a\1\ot a\2$
(Sweedler sigma notation with suppressed summation sign,
cf.\ Section~1.2 in~\cite{swe})

\item [$\de_{n}\;$] comultiplication applied $n$
times (Due to coassociativity we do not have
to remember where \de\ is put in consecutive
tensor products; $\de\sb na=a\1\ot\cdots\ot a\sb{(n+1)}\,$.)

\item [$\eta_{Y}\;\;$] unit map of an algebra $Y$ (often suppressed)

\item [$\epsilon\;\;\;\,$] counit (unless otherwise
obvious from the context)

\item [$S\;\;\,$ ] antipode, i.e.~$a\1 S(a\2)=\epsilon (a)=S(a\1)a\2\,$

\item [$A\sp{op}\;$] algebra identical with algebra $A$ as a vector space
but with the multiplication defined by $m\sb{A\sp{op}}=m\sb A\ci\tau$

\item [$\rho_{R}\;\;$] right coaction: $\rho_{R}(x) = x\0\ot x\1 $
(Sweedler sigma notation for comodules with 
suppressed summation sign, cf.\ p.32--3 in~\cite{swe})
\item [$\dr\;\,$] right coaction on the
`total space' of a quantum principal bundle

\item [$\Delta_{\cal R}\;\,$] right coaction on a right-covariant
differential algebra of
the `total space' given by $\mbox{$\forall\, m\in{\Bbb N}:\;$}
\mbox{$\dsr
(p\sb 0dp\sb 1\cdots dp\sb m)$}=\mbox{$(p\sb 0)\0d(p\sb 1)\0\cdots
d(p\sb m)\0\ot (p\sb 0)\1\cdots (p\sb m)\1$}$, where for
any \mbox{$n\in\{ 0,\cdots ,m\}$}, \mbox{$(p\sb n)\0\te (p\sb n)\1
=\dr p\sb n$} (A differential algebra is right-covariant iff
 \dsr\ determined by
the above formula is well-defined; cf.\ (21) and Section~4.2 in~\bmq .)

\item [$ad_{R}\;$] $:= (id\ot m)\ci (id\ot S\ot id)\ci
(\tau\ot id)\ci\de_{2}$ (right adjoint coaction;
$ad\sb Ra=a\2\te S(\! a\1\! )a\3\,$)

\item [$P\sp{co A}$] the space of right coinvariants ($P\sp{co A}:=\{p\in P |
\dr p=p\te 1\}$)

\item [$*_{\rho}\;\;\,$] convolution: $\forall\, f\!\in\!
\mbox{Hom}\sb k(Q,X),
g\!\in\! \mbox{Hom}\sb k(A,Y): f*_{\rho}g
= \m\ci (f\te g)\ci\rho_{R}$, where $(Q,\rho\sb R)$ is a right
$A$-comodule and $\m :X\te Y\ra Z$ is a multiplication map
(If $\rho_{R}$ equals $\dsr$, \dr\  or \de ,
we will use $*_{\cal R}$, $*_{R}$ or $*$ respectively
to denote the corresponding convolution.)

\item [$f\sp{-1}\;$] unless otherwise obvious from
the context, convolution inverse of $f$, i.e.~ \\
 $f\sp{-1}(a\1)f(a\2)=(f\sp{-1}\!  *\! f)(a)
=\epsilon (a)=(f\! *\! f\sp{-1})(a)=f(a\1)f\sp{-1}(a\2)\,$\\
(In general, one has:
 $x\sb{(0)}\ot\cdots\ot x\sb{(n)}\ot
f(x\sb{(n+1)})g(x\sb{(n+2)})\ot
x\sb{(n+3)}\ot\cdots\ot x\sb{(m)}\\
=x\sb{(0)}\ot\cdots\ot x\sb{(n)}\ot
(f*g)(x\sb{(n+1)})\ot x\sb{(n+2)}\ot\cdots\ot
x\sb{(m-1)}\;$ and\\
${\cal F}(x\0)g(x\1)\ot x\2\ot\cdots\ot x\sb{(m)}
=({\cal F}*\sb\rho g)(x\0)\ot x\1\ot\cdots\ot x\sb{(m-1)}$.)

\ei

All algebras are assumed to be unital and associative.
Now, let us recall the basic notions and
constructions of \bmq\  necessary to
establish the language used in this paper.
\bde [4.9 \bmq ]\label{qpbdef}
Let $P$ be an algebra over a field $k$, $A$ a
Hopf algebra over the same field, $N_{P}
\inc\Omega\sp{1}\! P$ a $P$-bimodule
defining the first order differential calculus
\op , $M_{A} \inc\ker\,\epsilon$ an $ad\sb R$-invariant right
ideal defining the bicovariant differential calculus
\oa , and
\[
\dr : P\lra P\ot A
\]
an algebra homomorphism making $P$ a right $A$-comodule
 algebra.\\ Then $(P,A,\dr ,N_{P},M_{A})$ is   called a quantum
principal  bundle iff:

\be

\item $\can :P\ot P\ni t\stackrel{\mbox{\em\scriptsize def.}}{\longmapsto}
(m\te id)\cc (id\te\dr )\, t\in P\ot A$ is a surjection (freeness condition),

\item $\dsr (N_{P})\inc N_{P}\ot A$
(right covariance of the differential
structure),

\item $\can (N_{P})\inc P\ot M_{A}$
 (fundamental vector field compatibility condition),

\item $\ker\,\mbox{\em\sf T} \inc P\hO\sp1(P\sp{co A}) P$ (exactness condition), 
where 
$\hO\sp1(P\sp{co A}) := \Omega\sp 1 P\sp{co A}/(N\sb
 P\cap\Omega\sp 1 P\sp{co A})$
\[
\mbox{and $\;$ \em\sf T}: \op\ni{[\alpha ]}
_{N_{P}}\stackrel{\mbox{\em\scriptsize def.}}{\longmapsto}
\llp(id\te\pi_{A})\cc \can\lrp\,\alpha \in P\ot (\ker\,\epsilon\, /\, M_{A})
\]
 (the map $\pi\sb A :
\ker\,\epsilon\ra\ker\,\epsilon\, /\, M_{A}$
is the canonical projection, and
$\alpha\!\in\!\ker\, m\sb P$).
 \ee
\ede
For simplicity, as well as to emphasize the
analogy with the classical situation, a quantum
principal  bundle is often denoted by
$P(B,A)$, where $B:=P\sp{co A}$ is the `base space' of the bundle. 
The map {\sf T} (denoted by
$\widetilde{~}\sb{N\sb P}$
in \bmq ) can
be more explicitly described by the formula
\[
\hsp{31mm}\T (pdq)=pq\0\ot
[q\1 ]\sb{M\sb A} - pq\ot 1 .\hsp{44mm}
\mbox{(cf.\ (24) in \bmq )}
\]

\bre\label{snake}{\em
Let $\;\T\!\sb U:\mbox{Ker}\, m\sb P\ra P\ot\mbox{Ker}\,\he\;$ and
$\;\T\!\sb{NM}:N\sb P\ra P\ot M\sb A\;$ be the appropriate restrictions 
of \cnl . It is straightforward to check that the following
diagrams are commutative diagrams (of left \mbox{$P$-modules}) with exact rows
and columns:\\

\beq\label{d1}
\def\normalbaselines{\baselineskip24pt
\lineskip3pt \lineskiplimit3pt }
\def\mapright#1{\smash{
\mathop{\!\!\!{-\!\!\!}-\!\!\!\longrightarrow\!\!\!}
\limits^{#1}}}
\def\mapdown#1{\Big\downarrow
\rlap{$\vcenter{\hbox{$\scriptstyle#1$}}$}}
\matrix{0&\mapright{}&\mbox{Ker}\,\T\!\sb{U}&\mapright{}&\mbox{Ker}\,\T\!\sb R
&\mapright{}&0&&&\phantom{.}\cr
&&\mapdown{}&&\mapdown{}&&\mapdown{}\cr
0&\mapright{}&\mbox{Ker}\, m\sb P&\mapright{}
&P\ot P&\mapright{m\sb P}&P&\mapright{}&0\cr
&&\mapdown{\T\!\sb{U}}&&\mapdown{\T\!\sb R}&&\mapdown{id}\cr
0&\mapright{}&P\ot\mbox{Ker}\,\he
&\mapright{}&P\ot A&\mapright{m\sb P\ci (id\ot\he)}
&P&\mapright{}&0\cr
&&\mapdown{}&&\mapdown{}&&\mapdown{}\cr
&&\mbox{Coker}\,\T\!\sb{U}&\mapright{}&\mbox{Coker}\,\T\!\sb R
&\mapright{}&0&\mapright{}&0\cr
}
\eeq\ \\

\beq\label{d2}
\def\normalbaselines{\baselineskip24pt
\lineskip3pt \lineskiplimit3pt }
\def\mapright#1{\smash{
\mathop{\!\!\!{-\!\!\!}-\!\!\!\longrightarrow\!\!\!}\limits^{#1}}}
\def\mapdown#1{\Big\downarrow
\rlap{$\vcenter{\hbox{$\scriptstyle#1$}}$}}
\matrix{0&\mapright{}&\mbox{Ker}\,\T\!\sb{NM}&\mapright{}&\mbox{Ker}\,\T\!\sb U
&\mapright{}&\mbox{Ker}\,\T\cr
&&\mapdown{}&&\mapdown{}&&\mapdown{}\cr
0&\mapright{}&N\sb P&\mapright{}
&\Omega\sp 1P&\mapright{\pi\sb P}&\op&\mapright{}&0\cr
&&\mapdown{\T\!\sb{NM}}&&\mapdown{\T\!\sb U}&&\mapdown{\T}\cr
0&\mapright{}&P\ot M\sb A&\mapright{}&P\ot\mbox{Ker}\,\he&\mapright{id\ot\pi\sb A}
&P\ot(\mbox{Ker}\,\he\, /\, M\sb A)&\mapright{}&0\cr
&&\mapdown{}&&\mapdown{}&&\mapdown{}\cr
&&\mbox{Coker}\,\T\!\sb{NM}&\mapright{}&\mbox{Coker}\,\T\!\sb U
&\mapright{}&\mbox{Coker}\,\T&\mapright{}&0\cr
}
\eeq\ \\

Applying the Snake Lemma (e.g., see Section~1.2 in \cite{aki}) to both diagrams,
we obtain the following two exact sequences:
\beq\label{s1}
0\lra\mbox{Ker}\,\T\!\sb U\lra\mbox{Ker}\,\T\!\sb R\lra 0
\lra\mbox{Coker}\,\T\!\sb U\lra\mbox{Coker}\,\T\!\sb R\lra 0
\phantom{vhvhvjhvhvsccsj......}
\eeq
\beq\label{s2}
0\lra\mbox{Ker}\,\T\!\sb{NM}\lra\mbox{Ker}\,\T\!\sb U\lra\mbox{Ker}\,\T 
\lra\mbox{Coker}\,\T\!\sb{NM}\lra\mbox{Coker}\,\T\!\sb U\lra\mbox{Coker}\,\T\lra 0
\eeq\ \\
Observe that the freeness condition means exactly that $\mbox{Coker}\,\T\!\sb R=0$,
which, by (\ref{s1}), is equivalent to $\mbox{Coker}\,\T\!\sb U=0$ (see (33) in
\bmq ). Note also that $\mbox{Ker}\,\T\!\sb{NM}=N\sb P\cap\mbox{Ker}\,\T\!\sb U$.
}\hfill{$\Diamond$}\ere

\bco\label{dcor}
$\!$Let $(P,A,\dr ,N_{P},M_{A})$ be a quantum
principal  bundle. Then \mbox{$\can(\! N\sb P\! )\! =\! P\te M\sb A$} 
(cf.~Example~4.11 in \bmq\ and the discussion below it).
\eco

{\it Proof.} From the exactness condition, we know that 
$\mbox{Ker}\,\T=\pi\sb P(P\Omega\sp 1\! B.P)$. On the other hand, since
$P\Omega\sp 1\! B.P\inc\mbox{Ker}\,\T\!\sb U$ and the map
$\pi\sb U:\mbox{Ker}\,\T\!\sb U\ra\mbox{Ker}\,\T$ in (\ref{s2}) is a restriction
of $\pi\sb P$ to $\mbox{Ker}\,\T\!\sb U$, we can conclude that $\pi\sb U$ is
surjective. Consequently, by the freeness condition and the exactness of (\ref{s2}),
$\mbox{Coker}\,\T\!\sb{NM}=0$, i.e.~\cnl$(\! N\sb P\! )=P\te M\sb A$.
\epf\\

The above corollary makes the following definition of a trivial quantum principal
bundle equivalent to the definition proposed in Example~4.11 in \bmq .

\bde\label{trbu}
A quantum principal bundle $(P,A,\dr ,N\sb P,M\sb A)$ is called trivial iff there
exists a convolution invertible map
(trivialization) $\Phi\in \mbox{\em Hom}\sb k(A,P)$ such that
\beq\label{trcov}
\dr\ci\Phi = (\Phi\ot id)\ci\de
\eeq
(i.e.~$\Phi$ is right-covariant) and $\Phi (1) = 1$. In such a case, $P$ is also
called a crossed product or cleft extension (see p.273 in \cite{sch2}).
\ede
\bde [1.1 \cite{sch2}]\label{galois}
Let $P$ be a right $A$-comodule algebra and $B$ be the algebra of all right
coinvariants. The comodule $P$ is called an $A$-Galois extension iff the canonical
left \mbox{$P$-algebra} and right $A$-coalgebra map
\[
\mbox{\em\sf T}\!\sb B:=(m\sb P\te id)\ci(id\te\sb B\dr):
P\ot\sb B P\ni p\ot\sb Bq\longmapsto pq\0\ot q\1\in P\ot A
\]
is bijective. 
\ede
\bpr\label{diagram}{\em\footnote{
This proposition is implicitly proved in \cite{tbt} (see Lemma~3.2 and the text
above it). The diagrammatic proof presented here was created during the author's
discussion with Markus Pflaum.}}
Let $P$, $A$ and $B$ be as above. A comodule algebra $P$ is an $A$-Galois extension
if and only if $P(B,A)$ is a quantum principal bundle with the universal calculus.
\epr
{\it Proof.} Consider the following commutative diagram (of left $P$-modules)
with exact rows and columns:\\

\beq\label{d3}
\def\normalbaselines{\baselineskip24pt
\lineskip3pt \lineskiplimit3pt }
\def\mapright#1{\smash{
\mathop{\!\!\!{-\!\!\!}-\!\!\!\longrightarrow\!\!\!}
\limits^{#1}}}
\def\mapdown#1{\Big\downarrow
\rlap{$\vcenter{\hbox{$\scriptstyle#1$}}$}}
\matrix{0&\mapright{}&P\hO\sp1\! B.P&\mapright{}&\mbox{Ker}\,\T\!\sb R
&\mapright{}&\mbox{Ker}\,\T\!\sb B&&&\phantom{.}\cr
&&\mapdown{}&&\mapdown{}&&\mapdown{}\cr
0&\mapright{}&P\hO\sp1\! B.P&\mapright{}
&P\ot P&\mapright{}&P\ot\sb B P&\mapright{}&0\cr
&&\mapdown{}&&\mapdown{\T\!\sb R}&&\mapdown{\T\!\sb B}\cr
0&\mapright{}&0
&\mapright{}&P\ot A&\mapright{id}
&P\ot A&\mapright{}&0\cr
&&\mapdown{}&&\mapdown{}&&\mapdown{}\cr
&&0&\mapright{}&\mbox{Coker}\,\T\!\sb R
&\mapright{}&\mbox{Coker}\,\T\!\sb B&\mapright{}&0\cr
}
\eeq\ \\

Again, we can apply the Snake Lemma to obtain an exact sequence
\beq\label{s3}
0\lra P\hO\sp1\! B.P\lra\mbox{Ker}\,\T\!\sb R\lra\mbox{Ker}\,\T\!\sb B\lra 0\lra
\mbox{Coker}\,\T\!\sb R\lra\mbox{Coker}\,\T\!\sb B\lra 0\; .
\eeq
Assume first that $P$ is an $A$-Galois extension. Then 
$\mbox{Ker}\,\T\!\sb B=0=\mbox{Coker}\,\T\!\sb B$, and, from the exactness of 
(\ref{s3}), we can infer that $\mbox{Coker}\,\T\!\sb R=0$ (freeness condition)
and $\mbox{Ker}\,\T\!\sb R=P\hO\sp1\! B.P\,$. 
On the other hand, by the exactness of (\ref{s1}),
we have $\mbox{Ker}\,\T\!\sb U=\mbox{Ker}\,\T\!\sb R\,$.  Hence
the exactness condition follows, and we can conclude that $P(B,A)$ is a quantum
principal bundle with the universal calculus.

Conversely, assume that $P(B,A)$ is a quantum
principal bundle with the universal calculus. Then 
$\mbox{Ker}\,\T\!\sb R=\mbox{Ker}\,\T\!\sb U=P\hO\sp1\! B.P$
and $\mbox{Coker}\,\T\!\sb R=0\,$. Consequently, again due to the exactness of
(\ref{s3}), we have that $\mbox{Ker}\,\T\!\sb B=0=\mbox{Coker}\,\T\!\sb B\,$,
i.e.~$P$ is an $A$-Galois extension.
\epf

\bde [\bmq ]\label{condef}
A left $P$-module projection $\Pi$ on \op\ is called a connection on\linebreak
 $P(B,A)$~iff
\be
\item $\ker\,\Pi =  \Omega\sp{1}_{hor}(P)$
($\,\mbox{\em Im}\,\Pi$ is called the space of vertical
 forms),
\item $\dsr\ci\Pi = (\Pi\ot id)\ci\dsr$ (right covariance).
\ee\ede 

Due to Proposition 4.10 in \bmq , a connection
form can be defined in the following way:

\bde \label{confordef}
A $k$-homomorphism $\omega : A\ra\op$ is called a
connection form on $P(B,A)$ iff it satisfies the
following properties:
\be
\item $\omega (k\oplus M_{A}) = 0$ (compatibility with
the differential structure),
\item $\mbox{\em\sf T}
\ci\omega = (id\ot\pi_{A})\ci \llp 1\ot
(id - \epsilon)\lrp$ (fundamental vector field condition),
\item $\dsr\ci\omega = (\omega\ot id)\ci
ad_{R}$ (right adjoint covariance).
\ee\ede

For every $P(B,A)$, there is a one-to-one
correspondence between connections and
connection forms. In particular, the connection
$\Pi\sp{\omega}$ associated to
a connection form $\omega$ is given by the
formula: $\Pi\sp{\omega}=m\sb{\Omega\sp 1(P)}\ci
(id\ot\omega )\ci\T$ ((47) in \bmq ).
Since $\Pi\sp{\omega}$
is a left $P$-module homomorphism, to calculate
$\Pi\sp{\omega}$ it suffices to
know its values on exact forms, and on exact forms
(47)\bmq\ simplifies to
\beq\label{confor}
\Pi\sp\omega\ci d = id*_{R}\omega .
\eeq

The following four definitions are based on Appendix~A\bmq .

\bde
Let $\Omega (P)$ be any differential algebra
having \op\ as in Definition~\ref{qpbdef}. For all $n\in\Bbb{N}$, the space 
$\llp P\ob P\lrp\sp{n}$ is called the space of horizontal
$n$-forms and is denoted by $\Omega\sp{n}_{hor}(P)$.
The space of horizontal \mbox{0-forms} is identified with~$P$.
\ede

\bde
Let $\Omega (B)$ be the differential algebra obtained by the restriction of
$\Omega (P)$. For all $n\in\Bbb{N}$, the space $\Omega\sp{n}\!
(\! B)P$ is called the
space of strongly horizontal $n$-forms and is denoted
by $\Omega\sp{n}_{shor}(P)$. The space of strongly horizontal \mbox{0-forms}
is identified with $P$.
\ede

Note that in the classical case $\Omega\sp{*}_{hor}(P)$
and $\Omega\sp{*}_{shor}(P)$ coincide.

\bde\label{tfdef}
 Let $\, (V,\rho_{R})\,$ be a right $A\sp{op}$-comodule algebra 
(see Remark~\ref{bialgebra}). Then\linebreak 
\mbox{$\phi\in \mbox{\em Hom}_{k}\llp V,\Omega (P)\lrp$}
 is called a pseudotensorial form on $P$ iff
\[
\dsr\ci\phi = (\phi\ot id)\ci\rho_{R}\, .
\]
A pseudotensorial form taking values in $\Omega\sp{*}_{hor}(P)$ 
(in $\Omega\sp{*}_{shor}(P)$) is called a tensorial (strongly tensorial)
 form on $P$. The space of all pseudotensorial,
tensorial and strongly tensorial \mbox{$n$-forms} ($n\geq 0$) will be denoted by
 $PT_{\rho}\llp V,\Omega\sp{n}\! (\! P)\lrp$,
$T_{\rho}\llp V,\Omega\sp{n}\! (\! P)\lrp$ and
 $ST_{\rho}\llp V,\Omega\sp{n}\! (\! P)\lrp$ respectively.
\ede

\bde [(68)\bmq ]\label{Ddef}
Let $\Pi$ be a connection on $P$. The $k$-homomorphism
$D$ from $\Omega\sp{*}\! P$ to $\Omega_{hor}\sp{*+1}\! P$
given by
\beq\label{689}
D : p\sb{0}dp\sb{1}\cdots dp\sb{n}\longmapsto
(id-\Pi )(dp\sb{0})\cdots (id-\Pi )(dp\sb{n})\, ,
\eeq
where $n\geq 0\,$, is called the exterior covariant derivative associated to~$\Pi$.
\ede 

To complete this vocabulary review, we recall some basic definitions used
in the Yang--Mills theory on projective modules. We choose here right rather
than left modules, but one should bear in mind that the formulation of this
formalism for left modules is analogous.

\bpr[cf.~p.369 in \cite{ms}]\label{spera}
Let $\cal B$ be an associative unital algebra over a commutative ring $k$.
Let $\cal L\sb k$ be a $k$-Lie subalgebra of the space of all $k$-derivations of
$\cal B$, and let $\cal E$ be any right $\cal B$-module admitting a connection. 
If $\hO(\! \cal B)$ is a differential graded subalgebra of 
\mbox{$\cal B\oplus\bigoplus\sb{n\geq 1}\sp\infty                         
\mbox{\em Hom}\sb k(\mbox{\large$\wedge$}\sp n\cal L\sb k,\, \cal B)$} 
with the differential given by (see the first section in~\cite{dkm}):
\bea
(d\ha)(X\sb 0,X\sb 1,\cdots ,X\sb n)\!\!\!
&=&\!\!\!\!\sum\sb{0\leq i\leq n}(-)\sp iX\sb i\ha
(X\sb 0,\cdots,X\sb{i-1},X\sb{i+1},\cdots ,X\sb n)\\
&+&\!\!\!\!\!\!\sum\sb{0\leq r<s\leq n}\!\!\! (-)\sp{r+s}\ha
([X\sb r,X\sb s],X\sb 0,\cdots,X\sb{r-1},X\sb{r+1},
\cdots ,X\sb{s-1},X\sb{s+1},\cdots ,X\sb n)\, ,
\eea
then 
\beq\label{nabla2}
\mbox{\large$\forall$}\,\xi\in\cal E,\; X,Y\in\cal L\sb k:\;\;
(\nabla\sp 2\xi)(X,Y)=
\left([\nabla\sb X,\nabla\sb Y]-\nabla\sb{[X,Y]}\right)(\xi)\, ,
\eeq
where, as in the classical differential geometry,
 $\nabla\!\sb Z\,\xi$ denotes $(\nabla\xi)(Z)$.
\epr

{\it Proof.} Straightforward.\hfill{$\Box$}

\bde[\cite{conri,ri}]
Let $\cal L$ be a finite dimensional Lie subalgebra
of $\,\mbox{\em Der}\, B$, let\linebreak 
$\{X\sb l\}\sb{l\in\{ 1,...,\mbox{\em\scriptsize dim}\,\cal L\}}$ 
be a basis of $\cal L$, and let $\hO(\! B)$ be 
a differential algebra defined as in Proposition~\ref{spera}. A~bilinear form 
\[
\{\;\, ,\;\}:\llp\mbox{\em End}\sb B(\cal E)\ot\sb B\hO\sp 2(\! B)\lrp
\times
\llp\mbox{\em End}\sb B(\cal E)\ot\sb B\hO\sp 2(\! B)\lrp
\lra
\mbox{\em End}\sb B(\cal E)
\]
is defined by
\[
\{\psi\sb 1,\psi\sb 2\}=\sum\sb{i<j}
\psi\sb 1(X\sb i\wedge X\sb j)\psi\sb 2(X\sb i\wedge X\sb j)\, .
\]\ede\ \vspace*{-2mm}

\bde[cf.~p.553--4 in \cite{conbook}]\label{pmdef}
Let $\cal E$ be a finitely generated projective right \linebreak
\mbox{$B$-module}.\\
\vspace*{-2mm}

1. A linear map 
$\nabla : \cal E\ot\sb B\Omega\sp *\! (\! B)\ra
\cal E\ot\sb B\Omega\sp{*+1}\! (\! B)$ is called a connection on $\cal E$ iff
\[
{\Lall}\;\xi\in\cal E,\, \ha\in\Omega(\! B)\, :\; 
\nabla(\xi\ot\sb B\ha )=(\nabla\xi )\ha +\xi\ot\sb Bd\ha 
\]\ \vspace*{-3mm}

2. The endomorphism 
$\nabla\sp 2\in\mbox{\em End}\sb{\Omega(B)}\llp\cal E\ot\sb B\hO (\! B)\lrp$
is called the curvature of a connection $\nabla$.
\\ \vspace*{-2mm}

3. If $B$ is a $*$-algebra and $\cal E$ is equipped with  a Hermitian metric
$\langle\;\, ,\;\rangle :\cal E\!\times\!\cal E\ra B\,$, then we say that a 
connection on $\cal E$ is compatible with this Hermitian metric iff
\beq\label{metcom}
{\Lall}\;\xi ,\eta\in\cal E\, :\; 
d\,\langle\xi ,\eta\rangle =
\langle\nabla\xi ,\eta\rangle + \langle\xi ,\nabla\eta\rangle
\eeq\ede\  \vspace*{-5mm}

\bre\label{ua}\em
The group $U(\cal E):=\{ U\!\in\!\mbox{End}\sb B(\cal E)\,|
\;\forall\,\xi ,\eta\!\in\!\cal E:\langle U\xi,U\eta\rangle =
\langle\xi ,\eta\rangle\}$
 of unitary automorphisms of $\cal E$ acts on the space of connections in the
following way: \mbox{$\nabla\to U\nabla U\sp*$}. This action maps compatible
connections to compatible connections (see near the end of Section~1
in \cite{conri}).\hfill{$\Diamond$}
\ere

\bre\em
The sign in the formula (\ref{metcom}) depends on whether we want
$d(a\sp*)=(da)\sp*$ or $d(a\sp*)=-(da)\sp*$. We have `+' in (\ref{metcom})
because here we choose that $d$ commute with $\sp*$.\hfill{$\Diamond$}
\ere

\bde[\cite{conri,ri}]
A trace $\cal T\sb E:\mbox{\em End}\sb B(\cal E)\lra k$ is given by
$\cal T\sb E(\xi\langle\zeta , .\rangle)=\cal T\sb B\langle\zeta ,\xi\rangle$,
where $\cal T\!\sb B$ is a trace on $B$.
\ede

\bco\label{traceco}
Let $\cal E=B\sp n$ for some $n\in\Bbb N$, and $N\in\mbox{\em End}\sb B(B\sp n)
=M\sb n(\! B)$. Then 
\[
\cal T\sb E(\! N)=(\cal T\sb B\ci Tr)(N)\, ,
\]
where $Tr$ is the usual matrix trace.
\eco

{\it Proof.} Straightforward. \hfill{$\Box$}

\bde[\cite{conri,ri}, cf.~\cite{conbook,dvkmm,dkm}]\label{ymfdef}
Let $\nabla$ be a connection on $\cal E$. The function $YM$ given by the formula
\beq\label{ymf}
YM(\nabla)=-\cal T\sb E\{\Theta\sb\nabla ,\Theta\sb\nabla\}\, , 
\eeq
where 
$\Theta$ is defined by
$\Theta\sb\nabla(X,Y)=[\nabla\sb X,\nabla\sb Y]-\nabla\sb{[X,Y]}$, is called
the Yang--Mills action functional.
\ede

\bex\label{group}\em
Let $G$ be a compact Lie group and $B=C\sp\infty(G)$. Let $W$ be 
a finite dimensional
vector space, $X(G,GL(W))$ a principal bundle, $\cal E=\hG\llp X(G,GL(W))
\times\sb{id}W\lrp$ the projective module of the smooth sections
of the $id$-associated vector bundle, and $g$ a Riemannian
metric on~$G$. (The choice of a metric on $\cal E$ makes no difference.)
Then, for any connection form $\omega$ on
$X(G,GL(W))$, one has $YM(\nabla\sp{\omega})=
-\int\sb G Tr (F\sp{\omega}\wedge\star F\sp{\omega})\,$,
where $\nabla\sp{\omega}$ is the covariant derivative (connection
on $\cal E$) associated to $\omega$, the symbol $\star$ denotes
the Hodge star associated to $g$, and $F\sp{\omega}$ is the
curvature 2-form of $\ho$ (see pages 336, 337 and 360 in~\cite{booss}).
\eex

\section{Strong Connections}
 
We can now proceed to formulate the notion of a strong
connection. This notion will be used to
justify the definition
of a global curvature form. It will also be needed for characterizing these
connections on a trivial $U\sb q(2)$-bundle that correspond to the hermitian 
connections on the free module associated with this bundle.
\bde\label{scdef}
Let $(P,A,\dr ,N_{P},M_{A})$ be a quantum
principal  bundle. A connection
$\Pi$ on $P$ is called strong iff $(id - \Pi )
(dP)\inc\Omega\sp{1}_{shor}(P)$ (see also
 Remark~\ref{scdefrem}).
\ede

For every connection $\Pi$, the left $P$-module homomorphism
$(id - \Pi )$ maps exact \mbox{1-forms} to horizontal, but
not necessarily to strongly horizontal, \mbox{1-forms}.
A strong connection $\Pi$ is defined by requiring that \mbox{$(id - \Pi )$}
sends exact \mbox{1-forms} to strongly horizontal \mbox{1-forms}.
It turns out that, for the trivial quantum principal 
bundles, the strongness of a connection means that the connection is induced 
from the base space of a bundle. (A~similar fact is described in Lemma~6.11
in~\cite{md3}.) More precisely, let $\hb\in\mbox{Hom}\sb k\llp A\, ,\, \ob\lrp$
be the map given by the formula
\beq\label{betadef}
\beta = \Phi *\omega *\Phi\sp{-1} + \Phi *(d\ci\Phi\sp{-1}).
\eeq
(This formula
can be obtained
by solving formula (37)\bmq\ for $\beta$
and extending the solution to a general
differential calculus.) It is straightforward to check that in the classical case
the thus defined \hb\ corresponds exactly to the pullback of a connection 1-form
with respect to the section associated with a given trivialization. Therefore, we
can think of \hb\ as a noncommutative analog of the aforementioned pullback of a
connection form (see also Remark~\ref{secrem}).
 (For a discussion of connection forms
which can be understood as elements of $\mbox{Hom}\sb k\llp A/k\, ,\, \ob\lrp$,
and which are also called quantum group gauge fields,
see Section~3 in \bmq .) With the help of (\ref{betadef}), we can characterize the
class of strong connections on any trivial quantum principal bundle in the following
way:

\bpr\label{scprop}
Let $(P,A,\dr ,N_{P},M_{A})$ be a trivial
quantum principal  bundle with a
trivialization $\Phi$ and a connection form
$\omega\,$, and let \hb\
 be as above.
Then $\,\Pi\sp{\omega}$ is strong if and only if $\beta (A)\inc\ob$.
\epr
{\it Proof.} It is known (e.g., see (27) in \cite{bmq}) that
\[
\mbox{\large$\forall$}\, p\!\in\! P\;\mbox{{\large$\exists$} 
$\sum_{i}$}b_{i}\te a_{i}\in B\te A\, :\; 
p = \mbox{$\sum_{i}$} b_{i}\Phi (a_{i})\, . 
\]
Using this fact, formulas (\ref{confor}), (\ref{trcov}), (37)\bmq , and
 the Leibniz rule it can be calculated that
\beq\label{Dp}
(id - \Pi\sp{\omega})(dp) = 
\sum_{i}\llp db_{i}.\Phi (a_{i})-b_{i}(\beta *\Phi )(a_{i})\lrp .
\eeq
(This calculation appeared in the preliminary
version of \bmq .) Clearly, $\beta (A)\inc\ob$
implies that $\Pi\sp{\omega}$ is a strong connection.
To prove the reverse implication, we will use the
following lemmas and corollary:
\ble\label{r-inv}
Let $\{ (Q\sb i,\rho\sb i)\}\sb{i\in I}$ and
$({\cal C},\rho\sb{\cal C})$ be right
$A$-comodules, and $\{ V\sb{ij}\}\sb{i,j\in I}$
be vector spaces, where $I$ is a
non-empty finite set. Assume also that, for all $i,j\in I$,
we have multiplication maps
\bea
&& \m\sb{ij}:Q\sb i\ot Q\sb j\lra V\sb{ij}\\
&& \m\sb{ij,l}:V\sb{ij}\ot Q\sb l\lra {\cal C}\\
&& \m\sb{i,jl}:Q\sb i\ot V\sb{jl}\lra {\cal C}
\eea
satisfying the associativity condition
\[
\forall\, i,j,l\in I: \m\sb{ij,l}\ci(\m\sb{ij}\ot id)= \m\sb{i,jl}\ci
(id\ot\m\sb{jl})\, ,
\]
and that coactions $\{\rho\sb i,\rho\sb{\cal C}\}\sb{i\in I}$
 are compatible in the following sense:
\[
\mbox{\large\boldmath$\forall$}\, i,j,l\in I\bc\,
q\sb i\in Q\sb i\bc\,
 q\sb j\in Q\sb j\bc\, q\sb l\in Q\sb l:
\rho\sb{\cal C}(q\sb iq\sb jq\sb l)=\rho\sb i(q\sb i)\rho\sb j(q\sb j)
\rho\sb l(q\sb l)\, .
\]
Then, for any $i,j,l\in I$, if $\kappa\sb i\in
 \mbox{\em Hom}\sb k(A,Q\sb i)$, $\varpi\sb j\in
\mbox{\em Hom}\sb
 k(A,Q\sb j)$ and $\lambda\sb l\in \mbox{\em Hom}\sb k(A,Q\sb l)$
are homomorphisms with the right-covariance properties:
\bea
&& \rho\sb i\ci\kappa\sb i=(\kappa\sb i\ot id)\ci\de\, ,\\
&& \rho\sb j\ci\varpi\sb j=(\varpi\sb j\ot id)\ci ad\sb R\, ,\\
&& \rho\sb l\ci\lambda\sb l=(\lambda\sb l\ot S)\ci\tau\ci\de\, ,
\eea
the homomorphism 
$\kappa\sb i*\varpi\sb j*\lambda\sb l$ is right-invariant, i.e.
\[
\rho\sb{\cal C}\ci (\kappa\sb i*\varpi\sb j*\lambda\sb l)=
(\kappa\sb i*\varpi\sb j*\lambda\sb l)\ot 1\, .
\]
\ele
{\it Proof.} A direct sigma notation computation
 proves this lemma.\hfill\mbox{$\Box$}

\bco\label{betainvcor}
Let $\omega$ be a connection form on a trivial
quantum principal bundle with a trivialization $\Phi$.
The map $\beta$
given by (\ref{betadef}) takes values in right-invariant
 differential forms, i.e.~
\[
\dsr\ci\beta = \beta\ot 1\, .
\]
\eco
{\it Proof.} Taking advantage of the formula
\[
\hsp{40mm}\dr\ci\Phi\sp{-1} = (\Phi\sp{-1}
\ot S)\ci\tau\ci\de\,
,\hsp{51mm}\mbox{((28) in \bmq )}\,
\]
one can deduce from Lemma~\ref{r-inv} that
\beq\label{1eq}
\dsr\ci (\Phi *\omega *\Phi\sp{-1})=
(\Phi *\omega *\Phi\sp{-1})\ot 1\, .
\eeq
(Observe that, since 
\[
 \llp(\Phi\sp{-1}\!\ot S)\ci\tau\ci\de\lrp
*(\dr\ci\Phi )=\eta\sb{P\te A}\ci\varepsilon =
(\dr\ci\Phi )*\llp
(\Phi\sp{-1}\!\ot S)\ci\tau\ci\de\lrp\, ,
\]
formula (28)\bmq\ follows by the same argument as used in the proof of
Proposition~\ref{gtprop}.)  Furthermore, as
\bea
&& \dsr\ci (d\ci\Phi\sp{-1})
\\ && =(d\ot id)\ci\dr\ci\Phi\sp{-1}
\hsp{18mm}
\mbox{(see the beginning of Section~1 for this property of \dsr )}
\\ &&
=\llp (d\ci\Phi\sp{-1})\ot S\lrp\ci\tau\ci\de\, ,
\eea
and $\eta\sb P\cc\epsilon$ is $ad\sb R$-covariant --- i.e.~
$\dr\ci\eta\sb P\ci\epsilon =\llp (\eta\sb P\ci\epsilon )\ot
id\lrp\ci
ad\sb R$ --- it also follows from  Lemma~\ref{r-inv} that
\beq\label{2eq}
\dsr\ci\llp\Phi *(\eta\sb P\ci\epsilon)*(d\ci\Phi\sp{-1})\lrp
=\llp\Phi *(\eta\sb P\ci\epsilon)*(d\ci\Phi\sp{-1})\lrp\ot 1\, .
\eeq
Combining formulas (\ref{1eq}) and (\ref{2eq}) we get the assertion
of the
corollary.\hfill\mbox{$\Box$}
\ble\label{rcovlem}
Let $(Q,\rho\sb 0)$ be a right $A$-comodule, and let
$\{ (Q\sb i,\rho\sb i)\}\sb
{i\in\{ 1,2\} }\,$, $V\sb{12}$ and $m\sb{12}$ be
as in
Lemma~\ref{r-inv}. If \mbox{$f\sb 1\in
 \mbox{\em Hom}\sb k(Q,Q\sb 1)$} is right-covariant
(i.e.~$\rho\sb 1\ci f\sb 1=(f\sb 1\ot id)\ci\rho\sb 0$) and
\mbox{$f\sb 2\in \mbox{\em Hom}\sb k(A,Q\sb 2)$}, then
\[
(id*\sb{\rho\sb 1}f\sb 2)\ci f\sb 1=f\sb 1*\sb{\rho\sb 0}f\sb 2\, .
\]
\ele
{\it Proof.} Straightforward. \hfill\mbox{$\Box$}
\bigskip

Assume now that
 $\Pi\sp{\omega}$ is strong.
Then, by setting $p=\Phi (a)$ in (\ref{Dp}),
we obtain:
\beq\label{bfeq}
(\beta *\Phi )(A)\inc\Omega\sp 1\sb{shor}(P)\, .
\eeq
Furthermore, it is a general fact that
\beq\label{shorlem}
(id *\sb{\cal R}\Phi\sp{-1})\llp\Omega
\sp{1}_{shor}(P)\lrp\inc\ob\, .
\eeq
Indeed, for any $b\in B$, $p\in P$,
\beq\label{idr}
 (id *\sb{\cal R}\Phi\sp{-1})(db.p)
=\llp m_{\Omega\sp{1}\! (\! P)}\ci
(id\ot\Phi\sp{-1})\lrp(db\0p\0
\ot b\sb{(1)}p\sb{(1)})
= db.\mbox{s}\sb\Phi (p),
\eeq
where
\beq\label{sphi}
\mbox{s}\sb\Phi :P\ni p\stackrel{\mbox{\scriptsize def.}}{\longmapsto}
(id*\sb R\Phi\sp{-1})(p)=p\0\Phi\sp{-1}(p\1)\in B\,
\eeq
is a left \mbox{$B$-module} homomorphism
that can be interpreted as the section
of $P(B,A)$ associated with
the trivialization $\Phi\,$ (see the subsequent remark, cf.\ Section~3.1 in
\cite{bk}). 
Taking again advantage of the formula (28)\bmq\
(see the proof of the Corollary~\ref{betainvcor})
we can infer that
\[
\dr\ci (id*\sb R\Phi\sp{-1})=(id*\sb R\Phi\sp{-1})\ot 1
\]
(cf.\ the second calculation in the proof
of Proposition~A.7 in \bmq ). Hence $\mbox{s}\sb\Phi$
indeed maps into~$B$, and (\ref{shorlem}) follows as claimed.
Combining (\ref{bfeq}) and (\ref{shorlem}) one can conclude that
\beq\label{incl}
\llp(id*\sb{\cal R}\Phi\sp{-1})\ci
(\beta *\Phi )\lrp(A)\inc\ob .
\eeq
On the other hand, taking into account Corollary~\ref{betainvcor},
one can see that $\beta *\Phi$ is \mbox{right-covariant:}
\bea
&& \hsp{19mm}\dsr\ci (\beta *\Phi )
\\ && \hsp{19mm}
=\llp (\dsr\ci\beta )*(\dr\ci\Phi )\lrp
\hsp{48mm}\mbox{(cf.\ Lemma~4.0.2
in \cite{swe})}
\\ && \hsp{19mm}
=m_{\Omega\sp{1}\! (\! P)\ot A}\ci\llp
 (\beta\ot 1)
\ot (\Phi\ot id)\lrp\ci\de_{2}
\\ && \hsp{19mm}
=\llp (\beta *\Phi)\ot id\lrp\ci\de\, .
\eea
Hence, by Lemma~\ref{rcovlem}, 
\[
(id*\sb{\cal R}\Phi\sp{-1})\ci
(\beta *\Phi )= (\beta *\Phi )*\Phi\sp{-1}= \beta\, ,
\]
and (\ref{incl}) reduces to $\beta
 (A)\inc\ob\,$, as needed.
\hfill{$\rule{7pt}{7pt}$}

\bre\label{secrem}\em
A natural question arises here as to whether
 or not one can, analogously to the
classical case, use $\,\mbox{s}\sb\Phi$ directly to compute the pullback of~\ho .
 For instance, in the case of the universal differential
calculus, we can define the pullback of
a differential \mbox{1-form} on $P$ with respect
to $\,\mbox{s}\sb\Phi$ to be
\[
\mbox{s}\sb\Phi\sp{*}\llp\sum_{i} p_{i}dq_{i}\lrp
= \sum_{i} \mbox{s}\sb\Phi (p_{i})(d\cc\mbox{s}\sb\Phi) (q_{i}).
\]
Clearly, since $\,\mbox{s}\sb\Phi \sp{*}(\Omega\sp{1}P)
\inc\Omega\sp{1}B$ and there exists a
non-strong connection on a trivial quantum principal bundle with the
universal differential calculus (see Example~\ref{z4}), Proposition~\ref{scprop}
 allows one to conclude that, in general,
$\,\mbox{s}\sb\Phi \sp{*}\ci\ho\,$ and $\beta$
given by (\ref{betadef}) do not coincide. (Otherwise \ho\ would always have to
be a strong connection form.) Furthermore, even if we assume that
$\omega$ is a strong connection form, the direct calculation of
$\,\mbox{s}\sb\Phi \sp{*}\ci\ho\,$ shows why, in the
noncommutative case, we cannot claim
$\,\mbox{s}\sb\Phi \sp{*}\ci\omega = \beta$. An advantage of defining $\beta$
by (\ref{betadef}) rather than by $\mbox{s}\sb\Phi\sp{*}\ci\omega = \beta$
 is that $\beta$ given by formula (\ref{betadef}) transforms
in a familiar manner under local gauge transformations (see Section~3 and (38)
in \cite{bmq}).

Let us also remark that, assuming the existence of $S\sp{-1}$,
one can define a quantum principal bundle section which is a right,
rather than left, \mbox{$B$-module} homomorphism from $P$ to $B$:
\[
\widetilde{\mbox{s}}\sb\Phi:=
m\sb P\ci\llp (\Phi\cc S\sp{-1})\ot id\lrp\ci\tau\ci\dr
\]
Much as in the case of $\,\mbox{s}\sb\Phi$,
it is straightforward to check
that $\,\widetilde{\mbox{s}}\sb\Phi$ is indeed a right \mbox{$B$-module}
homomorphism into $B$. From (28)\bmq\
(see the proof of Corollary~\ref{betainvcor}),
it is also clear that $\,\widetilde{\mbox{s}}\sb\Phi$
satisfies the equation $\,\widetilde{\mbox{s}}\sb\Phi\ci\Phi\sp{-1}=\he$.
This formula and an analogous formula $\,\mbox{s}\sb\Phi\ci\Phi=\epsilon$ for
$\,\mbox{s}\sb\Phi$ reflect the classical
geometry relationship between the section of a principal bundle and the
map from the total space to the structure group that are
associated to the same trivialization.
The pullback of a connection form \ho\ defined with the help of
$\,\widetilde{\mbox{s}}\sb\Phi$ has very similar
properties to the pullback of \ho\ defined with the help of
$\,\mbox{s}\sb\Phi$. Obviously,
$\,\mbox{s}\sb\Phi$ and $\widetilde{\mbox{s}}\sb\Phi$ coincide in the
classical case.
\hfill{$\Diamond$}\ere

Due to Proposition~\ref{scprop} and Proposition~4.6 \cite{bmq}
we know that every $\mbox{$\beta\in \mbox{Hom}\sb
k(A,\Omega\sp 1 \! B)$}$ vanishing on $k$ induces a strong connection
on a trivial quantum bundle with
the universal differential calculus (cf.~Section~6.4 in~\cite{md3}).
The quantum Dirac monopole considered in \bmq\ is an example of a strong
connection on a 
quantum bundle with a non-universal differential
calculus (for a proof of this fact see Corollary~6.4.4 in \cite{tbp}).
In what follows, we present an example of a weak (i.e.~non-strong) connection.
This example points out an interesting fact that the noncommutativity
alone of the `total space' $P$ or the `structure group' $A$ of a
quantum principal bundle
$P(B,A)$ cannot be held responsible
for the weakness of a connection.
Nor, for that matter, can we blame the
noncocommutativity of~$A$.

\bex \label{z4} \em
Let \zt\ denote the two-element group
represented by 0 and 2.
Also, let $P\! :=\!\M (\zf ,k)$ and $A\!
:=\!\M (\zt ,k)$ be the standard
commutative Hopf algebras over~$k$ (e.g., \
see \cite{mgtqg} or Section~2.2 in
\cite{abe}),
$\imath : \zt\hookrightarrow\zf$ be
the inclusion, and
\[
B:= \mbox{\large\{}b\in P\, |\,\Delta\sb R b: =
\llp(id\ot\imath\sp*)\cc\Delta\lrp (b) =
b\ot 1\mbox{\large\}}
\]
be the corresponding quantum homogeneous
space (see Section~5.1 in \cite{bmq}).
Since
\[
\kr\,\imath\sp*\inc
m\sb P\ci\llp (\kr\,\imath\sp*\cap B)\ot
P\lrp ,
\]
 by Lemma~5.2 \cite{bmq},
\mbox{$\llp P,A,(id\ot \imath\sp*)\ci\Delta
,0,0\lrp$}
is a quantum principal  bundle with the
universal differential calculus.
Now, let \mbox{$j:\zf\ra\zt$}
 be the surjection defined by $j(g)=\delta\sb{2,g}$.
Clearly, $\imath\ci j = id$ and
$j(h\sp{-1}gh)=h\sp{-1}j(g)h$ for all $g\!\in\!\zf\,
,\, h\!\in\!\zt\,$.
(Although \zt\ and \zf\ are additive groups,
to emphasize the usefulness of calculations shown in this example
even in more general cases, as well as to shorten some formulas,
we use the shorter and more abstract multiplicative notation.)
Hence, by Proposition~5.3 \cite{bmq}, we have the
canonical connection form given by the formula $\omega = (S*d)\ci j\sp*$.
Our task is now to show that the connection defined by $\omega$ is weak.
Suppose that this is not the case, i.e.~that $(id -\Pi\sp\omega)(dP)
\inc\Omega\sp 1\sb{shor} P\,$. Then, since it can be calculated that, 
for all $p\!\in\! P$,
\[
  (id - \Pi\sp\omega)(dp)
= d\,\Lblp p\sb{(1)}\, S
\Llp\!\llp\! (j\sp*\cc\imath\sp*)(p\2)\!\lrp\sb{(1)}\Lrp\!\Lbrp
 .\llp\! (j\sp*\cc\imath\sp *)(p\2)\!\lrp\2\, ,
\]
 and
$\Omega\sp 1\! P$ can be treated as the set of all
functions on $\zf\times\zf$
vanishing on the diagonal (e.g., \ see Section~2.6 in
\cite{coq}), we can conclude that, for any
$p\in P\bc\, g,r\in\zf\,\bc\,
 h\in \zt\,$,
\bea
&& \LAblp d\,\LAlp p\sb{(1)}\, S
\Llp\!\llp\! (j\sp*\cc\imath\sp*)(p\2)\!\lrp\sb{(1)}\Lrp\!\LArp
 .\llp\! (j\sp*\cc\imath\sp *)(p\2)\!\lrp\2\LAbrp (gh,r)\\
&&  =  \LAblp d\,\LAlp p\sb{(1)}\, S
\Llp\!\llp\! (j\sp*\cc\imath\sp*)(p\2)\!\lrp\sb{(1)}\Lrp\!\LArp
 .\llp\! (j\sp*\cc\imath\sp *)(p\2)\!\lrp\2\LAbrp (g,r)\, ,
\eea
\[
\;\;\;\;\;\;\mbox{i.e.~}\;\;
p(r) - p\llp ghj(h\sp{-1}g\sp{-1}r)\lrp = p(r)
 - p\llp gj(g\sp{-1}r)\lrp.
\]
 In particular, it implies that
\[
 \mbox{\large$\forall$}\, h\in\zt\, , g\in \zf\,
 :\; j(hg) = hj(g)
\]
(cf.~Lemma~5.5.5 in~\cite{tbp}). But $h=2$, $g=1$ do not verify that
equality and therefore we have a contradiction
proving that $\Pi\sp\omega$ is a weak connection.

Alternatively, since the pullback of the map
 \mbox{$\widetilde{\Phi}: \zf\ra\zt$} given by
\[
\widetilde{\Phi}:g\longmapsto\left\{
\begin{array}{ll}
0 & \mbox{for $g\leq 1$}\\
2 & \mbox{otherwise}
\end{array}
\right.
\]
is a trivialization of the quantum bundle $P(B,A)$ (see the Definition~\ref{trbu}),
one can prove that $\Pi\sp\omega$ is a weak connection by analyzing the map
$\beta$ associated to it (see (\ref{betadef})) and using
 Proposition~\ref{scprop}. Note also that the trivial connection
 (see Example~4.5 in \bmq ) induced by the
trivialization $\widetilde{\Phi}\,\sp *$ is, by Proposition~\ref{scprop}, strong.
\hfill{$\Diamond$}\eex

Our next example is concerned with a construction of both strong and non-strong
connections on certain quantum analogues of the $\Bbb Z\sb2$-Hopf fibration over the 
two-dimensional real projective space ($S\sp2\ra RP\sp2$).

\bex\label{rp2}\em
Let \pf\ be a unital free $*$-algebra over $\Bbb C$ generated by 
$\{ x\sb m\}\sb{m\in\{1,2,3\}}\,$,\linebreak 
(i.e.~\mbox{$P\sb F=\Bbb C<x\sb1,x\sb2,x\sb3,1>\,$}),
let $\cal I\sb0$ be a two-sided $*$-ideal of \pf\ generated by
\begin{eqnarray*}
&&\left\{x\sb m-x\sb m\sp*\, ,\;\mbox{$\sum\sb j$}a\sb jx\sb j\sp2-r\sp2\right\},\;
m,j\in\{1,2,3\}\, ,\; a\sb j,r\in\Bbb R\, ,\; a\sb j,r>0\, ,
\end{eqnarray*}
and let $\cal I\sb1$ be a two-sided $*$-ideal of \pf\ generated by
\begin{eqnarray*}
&&\mbox{\large\{}x\sb m-x\sb m\sp*\, ,\; x\sb1\sp2+x\sb2\sp2+
2\mbox{$\frac{1+q\sp4}{(1+q\sp2)\sp2}$}x\sb3\sp2-1\, ,\; 
x\sb1x\sb2-x\sb2x\sb1-2i\mbox{$\frac{1-q\sp4}{(1+q\sp2)\sp2}$}x\sb3\sp2\, ,\\ 
&&\ \; x\sb1x\sb3-\mbox{$\frac{q\sp{-2}+q\sp2}{2}$}x\sb3x\sb1-
i\mbox{$\frac{q\sp{2}-q\sp{-2}}{2}$}x\sb3x\sb2\, ,\;
x\sb2x\sb3-\mbox{$\frac{q\sp{-2}+q\sp2}{2}$}x\sb3x\sb2+
i\mbox{$\frac{q\sp{2}-q\sp{-2}}{2}$}x\sb3x\sb1\mbox{\large\}}\, ,\\
&&\ \; m\in\{1,2,3\}\, ,\;  q\in\Bbb R\, ,\; 1\geq |q|>0\, .
\end{eqnarray*}
(The second generator of $\cal I\sb1$ resembles the right-hand side of the formula
(164) in \cite{sm} that describes the metric on the $q$-Minkowski space discussed
in Section~7.2 of~\cite{sm}.)
The algebras $\pf /\cal I\sb0$ and $\pf /\cal I\sb1$ can be
regarded as noncommutative two-spheres. Indeed, $\pf /\cal I\sb0$ is 
`the most noncommutative two-sphere', and $\pf /\cal I\sb1$ corresponds to
the equator sphere ($c=\infty$) given by (7b) in \cite{po} 
(see Remark~\ref{equator}). Obviously, for $q=\pm 1$, the algebra 
$\pf /\cal I\sb1$ corresponds to the usual $S\sp2$. Since both $\pf /\cal I\sb0$
and $\pf /\cal I\sb1$ can be used in the same way to construct connections
on a noncommutative Hopf fibration, we denote, for the sake of brevity,
$\pf /\cal I\sb\nu$ by \pn , where $\nu\in\{0,1\}$. We also put
$x\sb{0j}=[x\sb j]\sb{\cal I\sb0}\,$, $x\sb{1j}=[x\sb j]\sb{\cal I\sb1}\,$,
\mbox{$a\sb{0j}=a\sb j$}, $j\in\{1,2,3\}\, ,\; 
a\sb{11}=1\, ,\, a\sb{12}=1\, ,\, a\sb{13}=2\frac{1+q\sp4}{(1+q\sp2)\sp2}\, ,\,
r\sb0=r\, ,\, r\sb1=1\,$, and thus define the coefficients 
$\{a\sb{\nu j}\, ,\, r\sb\nu\}\sb{\nu\in\{0,1\},\, j\in\{1,2,3\}}\,$.
Unless stated otherwise, all the following statements of this example will be
valid for any of the two values of $\nu$. The proposition below allows one
to turn \pn\ into a right $A$-comodule algebra, where 
$A=\mbox{Map}(\Bbb Z\sb2,\Bbb C)$.

\bpr\label{coaction}
Let \pf\ and $A$ be as above. Also, let \dr\ be a coaction of $A$ on \pf\ 
making it a right $A$-comodule algebra. If 
$\dr\ci *=(*\ot\overline{\phantom{a}})\ci\dr\, $, where $\overline{\phantom{a}}$
denotes the complex conjugation, and
\beq\label{co}
\dr :\pf\ni x\sb j\longmapsto x\sb j\ot (1-2\hd)\in\pf\te A\, ,
\eeq
where \hd\ is the map such that $\hd (-1)=1$ and $\hd (1)=0$, then 
$\dr (\cal I\sb\nu)\inc\cal I\sb\nu\te A$.
\epr
{\it Proof.} Clearly, $\dr (x\sb j-x\sb j\sp*)\in\cal I\sb\nu\te A\,$ for any
$j\in\{ 1,2,3\}$. Also, for any $j,l\in\{1,2,3\}$, we have 
\[
\dr (x\sb jx\sb l)=\dr (x\sb j)\dr (x\sb l)=
x\sb jx\sb l\ot (1-2\hd)\sp2=x\sb jx\sb l\te 1\, .
\]
Hence $\dr (\cal I\sb\nu)\inc\cal I\sb\nu\te A$, as claimed.
\epf\\

It follows now that a $*$-algebra homomorphism 
$\hD\sb\nu :\pn\ra\pn\te A$ given by the formula 
$\hD\sb\nu x\sb{\nu j}=x\sb{\nu j}\ot (1-2\hd)$ makes \pn\
a right $A$-comodule algebra.

\bpr\label{bundle}
Let \pn\  be a right $A$-comodule algebra as above. Then 
$(\pn ,A,\hD\sb\nu ,0,0)$ is a quantum principal bundle with the universal
differential calculus.
\epr
{\it Proof.} By Proposition~\ref{diagram}, it suffices to show that the
canonical map $\mbox{\sf T}\!\sb B:\pn\te\sb{B\sb\nu}\pn\ra\pn\te A$, where
$B\sb\nu:=\pn\sp{co A}$, is bijective.  First, however, let us prove the
following
\ble\label{bundlel}
Let $B\sb\nu$ be as above. Then $B\sb\nu$ is the space spanned by monomials
from \pn\ whose total degree is even, i.e.
\beq\label{bn}
B\sb\nu =\left\{\sum\sb{k\geq 1}\;\sum\sb{i\sb1,...,i\sb{2k}}
a\sb{i\sb1...i\sb{2k}}x\sb{\nu i\sb1}\cdots x\sb{\nu i\sb{2k}}\in\pn\, 
\mbox{\Large$|$}\; 
a\sb{i\sb1...i\sb{2k}}\in\Bbb C,\, i\sb1,...,i\sb{2k}\in\{1,2,3\}\right\}\, .
\eeq
\ele
{\it Proof.} To simplify notation, let us denote the right hand side of 
(\ref{bn}) by $\widetilde{B\sb\nu}$. Thanks to (\ref{co}), it is clear that
every element of $\widetilde{B\sb\nu}$ is right coinvariant. It is also clear
that every element of \pn\ can be written as 
$b\sb0+\sum\sb{j=1}\sp3b\sb jx\sb{\nu j}$ for some 
$\{b\sb l\}\sb{l\in\{0,...,3\}}\inc\widetilde{B\sb\nu}$. (Observe that any number
$c\in\Bbb C$ can be expressed as 
$c=cr\sb\nu\sp{-2}\sum\sb{i=1}\sp3a\sb{\nu i}x\sb{\nu i}\sp2\in
\widetilde{B\sb\nu}$.) Furthermore, since

\[
\hD\sb\nu p-p\te 1=b\sb0\te 1+\sum\sb{j=1}\sp3
(b\sb jx\sb{\nu j}\te 1-2b\sb jx\sb{\nu j}\te\hd )-b\sb0\te 1-\sum\sb{j=1}\sp3
b\sb jx\sb{\nu j}\te 1=-2\left(\sum\sb{j=1}\sp3b\sb jx\sb{\nu j}\right)\te\hd\, ,
\]

we can conclude that $\hD\sb\nu p=p\te 1\imp p=b\sb0\in\widetilde{B\sb\nu}\,$.
Hence $B\sb\nu=\widetilde{B\sb\nu}$, as claimed.
\epf\\

Now, consider a left \pn -module map 
$\widetilde{\T }\!\sb B :\pn\te A\ra\pn\te\sb{B\sb\nu}\pn $ given by the formula:

\[
\widetilde{\T }\!\sb B (1\te a)\longmapsto\left\{
\begin{array}{ll}
1\te\sb{B\sb\nu}1 & \mbox{for $a=1$}\\
\mbox{$\frac{1}{2}(1\te\sb{B\sb\nu}1-
r\sb\nu\sp{-2}\sum\sb{i=1}\sp3a\sb{\nu i}x\sb{\nu i}\ot\sb{B\sb\nu}x\sb{\nu i})$} 
& \mbox{for $a=\hd $}\, .
\end{array}
\right.
\]

Recall that every element of \pn\ can be written as 
$b\sb0+\sum\sb{j=1}\sp3b\sb jx\sb{\nu j}$ for some 
$\{b\sb l\}\sb{l\in\{0,...,3\}}\inc B\sb\nu$. Therefore, since
$\widetilde{\T }\!\sb B\ci\T\!\sb B$ is a left \pn -module map, it suffices to
check that 
\[
(\widetilde{\T }\!\sb B\cc\T\!\sb B)\llp 1\ot\sb{B\sb\nu} 
(b\sb0+\mbox{$\sum\sb{j=1}\sp3$}b\sb jx\sb{\nu j})\lrp=
1\ot\sb{B\sb\nu} (b\sb0+\mbox{$\sum\sb{j=1}\sp3$}b\sb jx\sb{\nu j})
\]
for arbitrary $\{b\sb l\}\sb{l\in\{0,...,3\}}\inc B\sb\nu$. 
With the help of Lemma~\ref{bundlel}, we have
\bea
&&
(\widetilde{\T }\!\sb B\cc\T\!\sb B)\llp 1\ot\sb{B\sb\nu} 
(b\sb0+\mbox{$\sum\sb{j=1}\sp3$}b\sb jx\sb{\nu j})\lrp
\\ &&
=b\sb0(\widetilde{\T }\!\sb B\cc\T\!\sb B)(1\ot\sb{B\sb\nu}1)
\mbox{$+\sum\sb{j=1}\sp3$}
b\sb j(\widetilde{\T }\!\sb B\cc\T\!\sb B)(1\ot\sb{B\sb\nu}x\sb{\nu j})
\\ &&
=b\sb0\ot\sb{B\sb\nu}1+\mbox{$\sum\sb{j=1}\sp3$}b\sb j
\widetilde{\T }\!\sb B(x\sb{\nu j}\te 1-2x\sb{\nu j}\te\hd )
\\ &&
=b\sb0\ot\sb{B\sb\nu}1+\mbox{$\sum\sb{j=1}\sp3$}
b\sb j\llp x\sb{\nu j}\ot\sb{B\sb\nu} 1
-2x\sb{\nu j}\widetilde{\T }\!\sb B(1\te\hd )\lrp
\\ &&
=b\sb0\ot\sb{B\sb\nu}1+\mbox{$\sum\sb{j=1}\sp3$}
\llp b\sb jx\sb{\nu j}\ot\sb{B\sb\nu}1-b\sb jx\sb{\nu j}\ot\sb{B\sb\nu}1
+r\sb\nu\sp{-2}\mbox{$\sum\sb{i=1}\sp3$}(a\sb{\nu i}b\sb jx\sb{\nu j}x\sb{\nu i}
\ot\sb{B\sb\nu}x\sb{\nu i})\lrp
\\ &&
=1\ot\sb{B\sb\nu}b\sb0+\mbox{$\sum\sb{j=1}\sp3$}\mbox{$\sum\sb{i=1}\sp3$}
r\sb\nu\sp{-2}a\sb{\nu i}\ot\sb{B\sb\nu}b\sb jx\sb{\nu j}x\sb{\nu i}\sp2
\\ &&
=1\ot\sb{B\sb\nu}(b\sb0+\mbox{$\sum\sb{j=1}\sp3$}b\sb jx\sb{\nu j}
\mbox{$\sum\sb{i=1}\sp3$}r\sb\nu\sp{-2}a\sb{\nu i}x\sb{\nu i}\sp2)
\\ &&
=1\ot\sb{B\sb\nu}(b\sb0+\mbox{$\sum\sb{j=1}\sp3$}b\sb jx\sb{\nu j})\, .
\eea
Thus we have shown that $\widetilde{\T }\!\sb B\ci\T\!\sb B=id$.
Furthermore, it is straightforward to verify that
$\T\!\sb B\ci\widetilde{\T }\!\sb B=id$. Hence $\widetilde{\T }\!\sb B$ is 
the inverse of $\T\!\sb B$ and consequently $\T\!\sb B$ is bijective,
as needed.
\hfill{$\rule{7pt}{7pt}$}

\bre\label{finite}\em
Note that, since $\mbox{Map}(\Bbb Z\sb2,\Bbb C)$ is finite dimensional,
the injectivity of $\T\!\sb B$ follows immediately from its surjectivity
and Theorem~1.3 in \cite{sch2}.
\hfill{$\Diamond$}\ere

\bre\label{equator}\em
Recall that a classical point of an algebra $B$ over a field $k$ is defined
as an algebra homomorphism from $B$ to $k$. For $q\neq\pm 1$, the space of all
classical points of $P\!\sb1$ is parameterized by all pairs $(x,y)\in\Bbb R\sp2$
subject to the relation $x\sp2+y\sp2=r\sb1=1$. Any such pair yields an algebra
homomorphism $f:P\!\sb1\ra\Bbb C$ via the formulas 
$f(x\sb{11})=x$, \mbox{$f(x\sb{12})=y$}, $f(x\sb{13})=0\,$. It is clear that the 
classical `subspace' of $P\!\sb1$ is precisely its equator. Hence the name 
`equator sphere'. (To see the correspondence between $P\!\sb1$ and the 
$C\sp*$-algebra defined by (7b)\cite{po}, put $\mu=q$ and 
\beq\label{corr}
x\sb{11}=\frac{i}{2}(B\sp*-B)\, ,\; x\sb{12}=-\frac{1}{2}(B\sp*+B)\, ,\;
x\sb{13}=\frac{-1-q\sp2}{2}A\, ;
\eeq
 cf.~the beginning of Section~7 in \cite{po}.)
Now, note that the quantum sphere employed in Section~5.2 of~\bmq\ 
($c=0$ in \cite{po})
can be, in the same manner, regarded as a north pole sphere. On the other hand,
the quantum principal bundles considered in Proposition~\ref{bundle} were 
constructed to generalize the usual Hopf fibration $S\sp2\ra RP\sp2$ 
(set $q=\pm 1$ in the bundle $(P\!\sb1,A,\hD\sb1,0,0)$) where
$\Bbb Z\sb2$ moves the points on the sphere to their antipodal counterparts
(see p.69 in \cite{tr}). 
On the north pole sphere used in \bmq , there is no other classical point 
to which the north pole could be moved under the free action of $\Bbb Z\sb2\,$.
This is why, in order to deform the 
Hopf fibration $S\sp2\ra RP\sp2$, we used here the equator sphere instead. 
\footnote{
I am grateful to Stanis\l aw Zakrzewski for explaining these things to me.}
\hfill{$\Diamond$}\ere

\bpr\label{sconnection}
Let $\pn (B\sb\nu ,A)$ be a quantum principal bundle as in 
Proposition~\ref{bundle} and Lemma~\ref{bundlel}. Also, let
$\ho\in\mbox{\em Hom}\sb{\Bbb C}(A,\hO\sp1\pn )$ be a homomorphism defined by
the formula
\[
\ho (a)=\left\{
\begin{array}{ll}
0 & \mbox{for $a=1$}\\
\mbox{$-\frac{1}{2r\sb\nu\sp2}\sum\sb{i=1}\sp3
a\sb{\nu i}x\sb{\nu i}dx\sb{\nu i}$} 
& \mbox{for $a=\hd $}\, .
\end{array}
\right.
\]
Then \ho\ is a strong connection form on $\pn (B\sb\nu ,A)$.
\epr
{\it Proof.} To prove that \ho\ is a connection form it suffices to verify that
$(\T\cc\ho )\hd =1\te\hd\ $. (Other conditions of Definition~\ref{confordef} are
immediately satisfied.) We have

\bea
(\T\cc\ho )\,\hd
&=&
-\frac{1}{2r\sb\nu\sp2}\sum\sb{i=1}\sp3
a\sb{\nu i}x\sb{\nu i}\T\!\sb R(1\te x\sb{\nu i}-x\sb{\nu i}\te 1)
\\ &=&
-\frac{1}{2r\sb\nu\sp2}\sum\sb{i=1}\sp3a\sb{\nu i}x\sb{\nu i}
(x\sb{\nu i}\te 1-2x\sb{\nu i}\te\hd\ -x\sb{\nu i}\te 1)
\\ &=&
1\ot\hd\, . \phantom{+\frac{1}{2}\ot 1}
\eea
Hence \ho\ is indeed a connection form. Our next step is to show that
\ho\ is strong (see Definition~\ref{scdef}). With the help of formula
(\ref{confor}) and the Leibniz rule, for any 
$\{b\sb l\}\sb{l\in\{0,...,3\}}\inc B\sb\nu\,$, we have
\bea
(\hP\sp\omega\cc d)(b\sb0+\mbox{$\sum\sb{j=1}\sp3$}b\sb jx\sb{\nu j})
&=&
-2\sum\sb{j=1}\sp3b\sb jx\sb{\nu j}\ho (\hd )
\\ &=&
r\sb\nu\sp{-2}\sum\sb{j=1}\sp3\sum\sb{i=1}\sp3a\sb{\nu i}b\sb jx\sb{\nu j}
x\sb{\nu i}dx\sb{\nu i}
\\ &=&
r\sb\nu\sp{-2}\sum\sb{j=1}\sp3\sum\sb{i=1}\sp3a\sb{\nu i}b\sb j
\llp d(x\sb{\nu j}x\sb{\nu i}\sp2)-d(x\sb{\nu j}x\sb{\nu i}).x\sb{\nu i}\lrp
\\ &=&
\sum\sb{j=1}\sp3b\sb jdx\sb{\nu j}-r\sb\nu\sp{-2}\!\!\!\sum\sb{i,j\in\{1,2,3\}}
\!\!\! a\sb{\nu i}b\sb jd(x\sb{\nu j}x\sb{\nu i}).x\sb{\nu i}\, .
\eea
Applying the Leibniz rule again, we obtain
\[
(id-\hP\sp\omega)\llp d(b\sb0+\mbox{$\sum\sb{j=1}\sp3$}b\sb jx\sb{\nu j})\lrp
=
db\sb0+\sum\sb{j=1}\sp3db\sb j.x\sb{\nu j}
+r\sb\nu\sp{-2}\!\!\!\sum\sb{i,j\in\{1,2,3\}}\!\!\!
a\sb{\nu i}b\sb jd(x\sb{\nu j}x\sb{\nu i}).x\sb{\nu i}\in\hO\sb{shor}\sp1\pn\, .
\]
Taking advantage of the fact that any $p\in\pn\ $ can be expressed as
$b\sb0+\mbox{$\sum\sb{j=1}\sp3$}b\sb jx\sb{\nu j}$ for some 
$\{b\sb l\}\sb{l\in\{0,...,3\}}\inc B\sb\nu\,$, we can conclude that \ho\
is strong.
\epf\\

\bpr\label{wconnection}
Let $\pn (B\sb\nu ,A)$ and \ho\ be as in the proposition above. A homomorphism
\linebreak\mbox{$\widetilde{\ho }\in\mbox{\em Hom}\sb{\Bbb C}(A,\hO\sp1\pn )$} 
 defined by
the formula
\[
\widetilde{\ho }(a)=\left\{
\begin{array}{ll}
0 & \mbox{for $a=1$}\\
\ho (\hd )+dx\sb{\nu l}\sp2 
& \mbox{for $a=\hd $}\, ,
\end{array}
\right.
\]
where $l\in\{1,2,3\}$, is a connection 1-form of a connection that is not strong.
\epr
{\it Proof.} Let $l$ be any fixed element of $\{1,2,3\}$. Since 
$\dsr(dx\sb{\nu l}\sp2)=dx\sb{\nu l}\sp2\ot 1$ and $\T(dx\sb{\nu l}\sp2)=0$,
it is clear that $\widetilde{\ho }$ is a connection 1-form. To prove that
$\hP\sp{\tilde{\omega}}$ is not a strong connection, we will demonstrate that
$(id-\hP\sp{\tilde{\omega}})(dx\sb{\nu l})
\;\; /\!\!\!\!\!\in\hO\sb{shor}\sp1\pn\, $.
With the help of formula (\ref{confor}), we have:
\[
(id-\hP\sp{\tilde{\omega}})(dx\sb{\nu l})=dx\sb{\nu l}+2x\sb{\nu l}
\widetilde{\ho }(\hd )=dx\sb{\nu l}+2x\sb{\nu l}\ho (\hd )
+2x\sb{\nu l}dx\sb{\nu l}\sp2=(id-\hP\sp\omega)(dx\sb{\nu l})
+2x\sb{\nu l}dx\sb{\nu l}\sp2
\]
Therefore, as $\hP\sp\omega$ is a strong connection, it is enough to show that
$x\sb{\nu l}dx\sb{\nu l}\sp2\;\; /\!\!\!\!\!\in\hO\sb{shor}\sp1\pn\, $.
To this end, let us put 
$B\sb\nu\sp c:=\left\{\sum\sb{j=1}\sp3b\sb j.x\sb{\nu j}\in\pn\ |\; 
b\sb j\in B\sb\nu\, ,\; j\in\{1,2,3\}\right\}$. Clearly, 
$\pn =B\sb\nu +B\sb\nu\sp c\,$. An argument similar to that used in the proof of
Lemma~\ref{bundlel} shows that $B\sb\nu\cap B\sb\nu\sp c=0$. 
It follows now that $\pn =B\sb\nu\oplus B\sb\nu\sp c\,$. As we consider here
the universal calculus on \pn , we have an isomorphism
$
\psi :\hO\sp1\pn\ni dp.u\to [p]\sb{\Bbb C}\ot u\in\pn /\Bbb C\ot\pn\, .
$
Furthermore, we have
\begin{eqnarray}\label{split}
\hO\sp1\pn 
&=&
\psi\sp{-1}\llp\psi (\hO\sp1\pn )\lrp 
\nonumber \\ &=&\psi\sp{-1}
(B\sb\nu /\Bbb C\ot\pn )\oplus\psi\sp{-1}
\llp (B\sb\nu\sp c\oplus\Bbb C)/\Bbb C\ot\pn\lrp
\nonumber \\ &=&
dB\sb\nu .\pn\oplus d(B\sb\nu\sp c\oplus\Bbb C).\pn 
\nonumber \\ &=&
\hO\sp1B\sb\nu .\pn\oplus d(B\sb\nu\sp c).\pn\, .
\end{eqnarray}
On the other hand, taking into account the isomorphism
$\hO\sp1\pn\cong\pn\ot (\pn /\Bbb C)$, one can show 
(with some help of the representation presented in Point~III.(a) of Proposition~4
in \cite{po} and formulas (\ref{corr})) that
$x\sb{\nu l}dx\sb{\nu l}\sp2\neq 0$. Therefore, since
$x\sb{\nu l}dx\sb{\nu l}\sp2=dx\sb{\nu l}\sp3-dx\sb{\nu l}.x\sb{\nu l}\sp2
\in d(B\sb\nu\sp c).\pn\,$, we can conclude, by virtue of (\ref{split}),
that $x\sb{\nu l}dx\sb{\nu l}\sp2\;\; /\!\!\!\!\!\in\hO\sb{shor}\sp1\pn\,$,
 as desired.
\epf\\

\bre\label{vertical}\em
The strong connection form \ho\ defined in Proposition~\ref{sconnection} is
non-trivial. Indeed, taking (\ref{confor}) into account, one can see that
\[
\hP\sp{\omega}(dx\sb{\nu l})=-2x\sb{\nu l}\ho (\hd )=r\sb\nu\sp{-2}
\sum\sb{i=1}\sp3a\sb{\nu i}x\sb{\nu l}x\sb{\nu i}dx\sb{\nu i}\, ,
\]
where, as before, $l$ is any fixed element of $\{1,2,3\}$.
On the other hand, it can be checked that 
$\left\{[x\sb{\nu i}]\sb{\Bbb C}\right\}\sb{i\in\{1,2,3\}}$
are linearly independent and that $x\sb{\nu l}\sp2\neq 0$. 
(Again, one can take advantage of the representation presented in Point~III.(a) 
of Proposition~4 in \cite{po} and formulas (\ref{corr}).) Hence, with the
help of an isomorphism $\hO\sp1\pn\cong\pn\ot (\pn /\Bbb C)$, it follows that
$\hP\sp\omega(dx\sb{\nu l})\neq 0$. Consequently, the space of vertical forms
(i.e.~Im $\hP\sp\omega$) is non-zero.
\hfill{$\Diamond\;\blacklozenge$}\ere 
\eex

We end our display of examples with a strict monoidal category
(see Section~6.1 in~\cite{shst}) example of a weak connection.

\bex\label{formal}\em
Similarly to Example~\ref{z4}, 
this example is concerned with a translation of the concept
of the canonical connection on a homogeneous space to a different set-up.
Only this time, the groups employed are neither
Abelian, nor finite. The former makes our bundle look more interesting,
the latter forces us to replace the algebraic tensor product by the
appropriate dual of the Cartesian product. More precisely, let
$\frak M$ be the image of the category of sets under the contravariant
functor $\mbox{Map}(\,\cdot\, ,k)$. The tensor product $\ot\sb{\frak M}$
defined by
\[
\mbox{Map}(X,k)\ot\sb{\frak M}
\mbox{Map}(Y,k)=\mbox{Map}(X\!\times\! Y,k)
\]
makes $\frak M$ a strict monoidal category. Moreover, if $X$ is a group,
then $\mbox{Map}(X,k)$  is an \mbox{$\frak M$-Hopf} algebra, where the
definition of an \mbox{$\frak M$-Hopf} algebra is the same as
that of a Hopf algebra but with the tensor product taken to be
$\ot\sb{\frak M}\,$. In what follows we define several gauge theoretic
notions in the setting of the category $\frak M\,$:
\be
\item $\Omega\sp 1\sb{\frak M}X:=\{
F\!\in\!\mbox{Map}(X\!\times\! X,k)\, |\,
\forall\, x\!\in\! X:F(x,x)=0\}$ (\mbox{$\mbox{Map}(X,k)$-bimodule} of
\mbox{$\frak M$-differential} \mbox{\mbox{1-forms}})

\item $\Delta\sb{\frak M-R}:=R\sp *$ (right coaction), where
$R:X\times G\ra X$ is a right free action of the group $G$ on $X$
(e.g., see p.55 in~\cite{tr}; $G$ will denote
a group and $R$ its right free action throughout the rest of this example)

\item $\Delta\sb{\frak M-{\cal R}}:\ox\ra\{
 K\!\in\!\mbox{Map}(X\!\times\! X
\!\times\! G,k)\; |\;\forall\; x\!\in\! X,\, g\!\in\!
 G:K(x,x,g)=0\}$ (right coaction
on \mbox{$\frak M$-differential} \mbox{\mbox{1-forms}}), where
$\mbox{\large$\forall$}\; F\!\in\!\ox\lc\,
x,y\!\in\! X\lc\, g\!\in\! G:
(\Delta\sb{\frak M-{\cal R}}F)(x,y,g):=F\llp R(x,g)\bc R(y,g)\lrp$

\item A triple $\llp\mbox{Map}(X,k)\lc\mbox{Map}(G,k)\lc
\Delta\sb{\frak M-R}\lrp$ is called an \mbox{$\frak M$-principal}
bundle and, for simplicity, denoted by $(X,G,R)$.

\item $\oxh :=\mbox{\large$\{$} F\!\in\!\ox \;
|\;\mbox{\large$\forall$}\; x\!\in\!
 X\lc\, g\!\in\! G: F\llp R(x,g)\lc\, x\lrp
=0\mbox{\large$\}$}$ (horizontal \mbox{$\frak M$-differential}
\mbox{\mbox{1-forms}})
\item $B:=\mbox{Map}(X/G,k)$ (base space of $(X,G,R)$)
\item $\oxs :=\mbox{\large$\{$} F\!\in\!\oxh\;
|\;\mbox{\large$\forall$}\; x,y\!\in\!
X\lc\, g\!\in\! G:
F\llp R(x,g)\lc\, y\lrp =F(x,y)\mbox{\large$\}$}$ (strongly
horizontal \mbox{$\frak M$-differential}
\mbox{\mbox{1-forms}})
\item Let $(X,G,R)$ be an  \mbox{$\frak M$-principal} bundle and let
\[
\widetilde\Pi :=\widetilde\Pi\sb
1\times\widetilde\Pi\sb 2
\in\mbox{$\mbox{Map}(X\!\times\! X,X\!\times\!
 X)$}
\]
 be an idempotent satisfying
the following conditions:
\bi
\item [a)] $\mbox{\large$\forall$}\; x,y\in
 X:\widetilde\Pi\sb 1(x,y)=x$,
\item [b)] $\mbox{\large$\forall$}\; F\in\ox :\;
F\ci\widetilde\Pi =0\iff F\in\oxh$,
\item [c)] $\mbox{\large$\forall$}\; x,y\!\in\! X\lc\, g\!\in\!
G: \widetilde\Pi\sb 2\llp
R(x,g)\,\lc\, R(y,g)\lrp =R\llp\widetilde\Pi
\sb 2(x,y)\,\lc\, g\lrp$.
\ei
Then  $\widetilde\Pi\sp *$ is called an
\mbox{$\frak M$-connection}
and is denoted by $\Pi\sb{\frak M}$.
\item An \mbox{$\frak M$-connection} $\Pi
\sb{\frak M}$ is called a strong
 \mbox{$\frak M$-connection} iff
\[
\mbox{\large\boldmath$\forall$}\,
 x,y\!\in\! X\lc\; g\!\in\! G:\;
\widetilde\Pi\sb 2\llp R(x,g)\,\lc\, y\lrp\\
 =
\widetilde\Pi\sb 2(x,y)\, .
\]
\ee
The above definitions were constructed so that the  \mbox{$\frak M$-objects}
thus defined become the corresponding
`quantum objects' (the universal differential calculus assumed)
when both $X$ and $G$ are finite and the  \mbox{$\frak M$-tensor} product
is the same as the algebraic tensor product. In particular,
one can rethink and equivalently
describe Example~\ref{z4} in  \mbox{$\frak M$-terms}.
 Doing so blurs the view of general principles making that example work,
but it allows one to have a better insight into its concrete mathematical fabric.

Clearly, if $H$ is a subgroup of $G$ acting
on $G$ on the right by the group
multiplication (let us denote this action by
$R\sb G$), then
$(G,H,R\sb G)$ is an  \mbox{$\frak M$-principal}
 bundle. Furthermore,
any surjection \mbox{$j:G\ra H$} satisfying
 $j\ci\imath = id$, where
\mbox{$\imath :H\hookrightarrow G$} is the
 inclusion, and
\beq\label{jad}
\forall\, g\in G,h\in H:\, j(h\sp {-1}gh)=h\sp
 {-1}j(g)h
\eeq
yields an \mbox{$\frak M$-connection} on
$(G,H,R\sb G)$. Indeed, let
$\widetilde\Pi\sp j(g,r):=\llp g,\,
gj(g\sp{-1}r)\lrp$, for any
$g,r\in G$. Then, for all $g,r\in G\bc\,
 h\in H$, we have:
\be
\item $(\widetilde\Pi\sp j)\sp 2(g,r)=\lblp g,\,
 gj\llp g\sp{-1}gj(g\sp{-1}r)\lrp\lbrp
=\llp g,\, gj(g\sp{-1}r)\lrp =\widetilde\Pi\sp j(g,r)$,
where the middle equality is implied by the formula $j\ci\imath =id$.

\item $\widetilde\Pi\sp j\sb 1(g,r)=g$ (obvious).

\item $F\in\Omega\sp 1\sb{\frak M-hor} G
\Rightarrow F\ci\widetilde\Pi\sb j=0$ (obvious).

\item For any $F\in\mbox{Map}( G\!\times\! G,k)$, the implication
\[
F\ci\widetilde\Pi\sp j=0\Rightarrow
F\in\Omega\sp 1\sb{\frak M-hor} G
\]
is a consequence of the fact that
$\widetilde\Pi\sb j$ restricted to
\mbox{$\bigcup\sb{g\in G}(g,gH )$}
coincides with the inclusion of
\mbox{$\bigcup\sb{g\in G}(g,gH )$} in $ G\times G$.

\item $\widetilde\Pi\sp j\sb 2(gh,rh)=
ghj(h\sp{-1}g\sp{-1}rh)=gj(g\sp{-1}r)h
=\widetilde\Pi\sp j\sb 2(g,r)h$, where
the middle step follows from
(\ref{jad}).
\ee
Hence, $(\widetilde\Pi\sp j)\sp *$ is an \mbox{$\frak M$-connection}.
Moreover, much as in Example~\ref{z4}, if
$(\widetilde\Pi\sp j)\sp *$ is a strong
\mbox{$\frak M$-connection}, then, for all $g\in G , h\in H$, $hj(g)=j(hg)$.

Next, we proceed to consider a special case of $G$ and $H$.
To do so, first we need a definition of an algebraic formal group:

\bde [cf.\ \cite{ha} and Appendix~A in \cite{mgtqg}]\label{afg}
Let $\frak g$ be a finite dimensional Lie algebra ($\dim \frak g=n$),
and let $\{ E\sp\nu\}\sb{\nu\in\{ 1,\cdots ,n\}}$
be a basis of $\frak g$.
Also, let $F$ be the formal group law
(see, e.g., Section~9.1 and Section~1.1 in~\cite{ha}) given by the
Baker--Campbell--Hausdorff formula determined by $\frak g$
and the basis $\{ E\sp\nu\}\sb{\nu\in\{ 1,\cdots ,n\}}$
(see Appendix~A in~\cite{mgtqg}), so that to every
pair of $n$-tuples of formal power series in a finite number of
variables with no free term we can assign another such $n$-tuple, i.e.
\bea
&&
F\lblp
\llp p\sb 1(t\sb 1,\cdots ,t\sb l)\lc\cdots\lc p\sb n(t\sb 1,\cdots
,t\sb l)\lrp\lbc
\llp q\sb 1(s\sb 1,\cdots ,s\sb m)\lc\cdots\lc q\sb n(s\sb 1,\cdots
,s\sb m)\lrp\lbrp  \\ &&
=\llp r\sb 1(t\sb 1,\cdots ,t\sb l,s\sb 1,\cdots ,s\sb m)\lc\cdots\lc
r\sb n(t\sb 1,\cdots ,t\sb l,s\sb 1,\cdots ,s\sb m)\lrp\, .
\eea
Symbolically, we will write \mbox{
$\llp\, p\sb 1(t\sb 1\, ,\,\cdots\, ,\, t\sb l)\,\lc\,\cdots\,\lc\,
 p\sb n(t\sb 1\, ,\,\cdots\, ,\, t\sb l)\,\lrp$}
as $\exp (p\sb\nu E\sp\nu )$ and \linebreak\mbox{$F\llp\exp
(p\sb\nu E\sp\nu )\lc\exp (q\sb\mu E\sp\mu )\lrp$}~as~\mbox{
$\exp (p\sb\nu E\sp\nu )\exp (p\sb\nu E\sp\nu )$}.
 (The Einstein convention of summation over repeating indices is
assumed here and throughout the rest of this example.)
The group generated with the use of $F$ by the $n$-tuples
$\{\exp (t\sb mX)\}\sb{m\in{\Bbb N},X\in\frak g}\,$, where
$\{ t\sb m\}\sb{m\in{\Bbb N}}$
are formal power series in one variable with all
but the linear coefficients vanishing, is called the algebraic formal group
associated to $\frak g$ and is denoted by $E\sb{\frak g}$.
\ede

Now, let $G=\esl$, $H=\esu$
and let \mbox{$j:\esl\ra\esu$} be the surjection defined by
$j\llp\exp (p\sb l E\sp l +p\sb\mu E\sp\mu)\lrp = \exp (p\sb l E\sp l)$,
where $\{E\sp l\}\sb{l\in\{ 1,2,3\}}$ is a fixed basis of ${\frak s\frak u}(2)$, 
and $\{ E\sp\mu\}\sb{\mu\in\{ 1,2,3\}}$ is a fixed basis of
$i{\frak s\frak u}(2)$. It is clear that 
$(\esl ,\esu ,R\sb{E\sb{\frak s\frak l(2,\Bbb C)}})$ is an
\mbox{$\frak M$-principal} bundle, and $j\ci\imath = id$.
Thus, to see that $j$ induces an
\mbox{$\frak M$-connection}, it suffices to prove the following:

\ble\label{lemex}
$\;\mbox{\large$\forall$}\, g\in \esl ,
 h\in \esu\, : \; j(h\sp{-1}gh)=h\sp{-1}j(g)h$
\ele
{\it Proof.} Since the formal power series
determining elements of $\esl$ are generated by the
Baker--Campbell--Hausdorff formula, we know that
\bea
\mbox{\large$\forall$}\, g\!\in\!\esl\;\,\mbox{\large$\exists$}
\,\epsilon\sb g>0,n\!\in\!{\Bbb N}\;\,\mbox{\large$\forall$}\,
\overline{x}\in\{ (z\sb 1,\cdots ,z\sb n)\in\Bbb R\sp n\, |
\; z\sb 1\sp 2+\cdots +z\sb n\sp 2<\epsilon\sb g\sp 2\}:&&
\\ 
 V\sb{\overline{x}}(g):=\exp\llp p\sb l
(\overline{x})E\sp l + p\sb\mu (\overline{x})E\sp
\mu\lrp\in SL(2,{\Bbb C})\, .&&
\eea
This means that the formal power series
$\{ p_{l},p_{\mu}\}\sb{l,\mu\in\{ 1,2,3\}}$
defining $g$ are convergent when evaluated at
 any $\overline{x}$ in an $\epsilon$-neighborhood
of $0\in {\Bbb R}\sp n$.
Now, let $g$ be an arbitrary element of $\esl$
and $h$ be any element of
$\esu$ of the form $h=\exp (tX)$, where 
\mbox{$X\!\in\!{\frak s\frak u}(2)$} and $t$
is understood as a formal power series in one variable. Also, let 
$\epsilon :=\min\{\epsilon\sb{h\sp{-1}j(g) h}\; ,\;
\epsilon\sb{j(h\sp{-1} gh)}\} $,
 $ \overline{y} :=(\overline{t} ,\overline{x})\in
{\Bbb R}\sp{n+1}$ and $\widetilde{j}$ be
the map defined on all elements of $SL(2,{\Bbb C})$ of the form
 $\exp(t\sb l E\sp l + t\sb\mu E\sp\mu)$, 
$\forall\, l,\mu :\, t\sb l,t\sb\mu\in {\Bbb R}, $ 
by the formula
 \[
\widetilde{j} \llp\exp (t\sb l
 E\sp l+t\sb\mu E\sp \mu )\lrp =\exp
(t\sb l E\sp l)\, .
\]
Then, using the fact that the splitting ${\frak s\frak l}(2,{\Bbb C}) = 
{\frak s\frak u}(2)\oplus i{\frak s\frak u}(2)$ is $ad$-invariant,
for every $\overline{y}$ of length smaller than $\epsilon$, we have
\bea
&& (V\sb{\overline{y}} \cc j)\,\llp\exp( -tX)\exp( p\sb l
 E\sp l+p\sb\mu E\sp\mu)\exp(tX)\lrp \\
&& = (\widetilde{j}\cc V\sb{\overline{y}})\,
\llp\exp(-tX)\exp(
p\sb lE\sp l +p\sb\mu E\sp\mu)\exp(tX)
\lrp \\
&& =\widetilde{j}\lblp\exp( -\overline{t}X)\exp
\llp p\sb l
 (\overline{x})E\sp l + p\sb\mu(
\overline{x})E\sp\mu\lrp\exp(\overline{t}X)\lbrp \\
&& =\widetilde{j}\lblp\exp\llp p\sb l
(\overline{x})
\exp(-\overline{t}X) E\sp l\exp(\overline{t}X) +
p\sb\mu(\overline{x}) \exp( -\overline{t}X)
 E\sp\mu
\exp(\overline{t}X)\lrp\lbrp \\
&& = \exp\llp p\sb l(\overline{x})\exp(-
\overline{t}X)
 E\sp l\exp(\overline{t}X)\lrp \\
&& = \exp(-\overline{t}X)\exp\llp p\sb
l(\overline{x})E\sp l\lrp\exp(\overline{t} X) \\
&& = V\sb{\overline{y}} \llp\exp( -tX)\exp( p\sb
l E\sp l)\exp(tX)\lrp \\
&& = V\sb{\overline{y}} \lblp\exp( -tX)
 j\llp\exp( p\sb l
E\sp l+p\sb\mu E\sp\mu)\lrp\exp(tX)
\lbrp .
\eea
Thus, the formal power series defining
$j(h\sp{-1} gh)$ and $h\sp{-1}j(g)h$
coincide when evaluated on an open
 neighborhood of $0\in{\Bbb R}\sp {n+1}$,
 and hence are identical. To end the proof,
we need to note that $ad:
\esu\ra\mbox{Aut}(\esl )$ is a
homomorphism and, since every element of
$\esu$ is generated by elements of the form
$\exp(tX)$, the formula
$j(h\sp{-1}gh) = h\sp{-1}j(g)h$ is 
valid for all $g\in\esl$ and $h\in\esu\,$, as claimed.
\hfill $\Box$
\bigskip

Finally, since $h=\exp (tY)$, $g=\exp(sZ)$,
 where $Y:=\mbox{\scriptsize $\pmatrix{0&1\cr -1&0}$}$, 
$Z:=\mbox{\scriptsize $\pmatrix{1&0\cr 0&-1}$}$, do not satisfy
$j(hg) = hj(g)$, the \mbox{$j$-induced}
\mbox{$\frak M$-connection} is non-strong, as
desired.\footnote{
I am very grateful to Philip Ryan 
for pointing out that $Y,Z$
provide the desired counterexample to the formula $j(hg) = hj(g)$.}

It seems proper to mention at this point
 that it would be interesting to see to what extent gauge theory on
quantum principal bundles can
work in some more interesting categories and whether 
the monoidal reconstruction (see
 Section~8.2 in \cite{shst}) can be
extended to reconstruct bundles and
connections.
\hfill{$\blacklozenge$}\eex

\section{Gauge Transformations}

The next natural step is to determine how
strong connections behave under quantum gauge transformations. To do so,
we must first define gauge transformations of a quantum principal bundle.
One can define the group of quantum
gauge transformations as the group of convolution-invertible elements
of $\mbox{Hom}\sb k(A,P)$ which intertwine
$\Delta\sb R$ with $ad\sb R$, and satisfy $f(1)=1$. (The same definition is
considered in Proposition~5.2 of~\cite{tbt}.)
Then one can define their action on connection forms in a way analogous
 to the action of their classical counterparts on the classical connection
 forms (see \cite{booss}). Quantum gauge
transformations defined in this manner
generalize locally defined quantum gauge transformations
discussed in Section~3 of~\bmq\ (cf.\ Section~3 in~\cite{md4}).

\bde\label{gtrans}
Let $P(B,A)$ be a quantum principal bundle.
A $k$-homomorphism \mbox{$f:A\ra P$} is called a gauge
transformation iff\vspace*{3mm}

\hsp{5mm} 1. $f$ is convolution invertible,

\hsp{5mm} 2. $\dr\ci f = (f\ot id)\ci ad\sb R,$

\hsp{5mm} 3. $f(1)=1$.
\ede

\bpr\label{gtprop}
The set of all gauge transformations of a
quantum principal  bundle
is a group with respect to convolution.
We denote this group by $GT(P)$.
\epr
{\it Proof.} A routine sigma notation calculation
verifies the following lemma:

\ble\label{adlem}
$\,$ Let $\,\{ (Q\sb i,\rho\sb i)\}\sb{i\in\{ 1,2\} }\,$,
$\,V\sb{12}\,$ and $\,m\sb{12}\,$ be as in Lemma~\ref{r-inv}.
Then, for all\linebreak
\mbox{$\mbox{\em f}\sb 1\in\mbox{\em Hom}\sb k(A,Q\sb 1)$},
\mbox{$\mbox{\em f}\sb 2
\in\mbox{\em Hom}\sb k(A,Q\sb 2)$},
\[
\llp (\mbox{\em f}\sb 1\te id)\ci ad\sb R\lrp *
\llp (\mbox{\em f}\sb 2\te id)\ci ad\sb R\lrp =
\llp (\mbox{\em f}\sb 1\! *\!\mbox{\em f}\sb 2)\ot
id\lrp\ci ad\sb R\, .
\]
\ele

Hence, since the map
\[
(\dr\ci ):\mbox{Hom}\sb k(A,P)\ni\mbox{f}\longmapsto\dr\ci\mbox{f}\in
\mbox{Hom}\sb k(A,P\te A)
\]
 is an algebra homomorphism (cf.\ Lemma~4.0.2 in \cite{swe}),
the set of all gauge
 transformations is closed
under the convolution. Furthermore,
it follows from the same reason that
\mbox{$\dr\ci\mbox{f}\,\sp{-1}=(\dr\ci\mbox{f}\, )\sp{-1}$}. Therefore,
as $f(1)=1$ implies $f\sp{-1}(1)=1$,
by putting $\mbox{f}\sb 1=f$ and $\mbox{f}\sb 2=f\sp{-1}$
in Lemma~\ref{adlem}, we can also conclude the existence of the inverse.
 \hfill $\Box$
\bigskip

When defined in this way,
quantum gauge transformations are unwilling to preserve the
property $\omega(M\sb A) = 0$ (see Definition~\ref{confordef} and
Proposition~\ref{gact}) defining a connection
form $\omega$ on a bundle with a general differential calculus. This is the case
even if one assumes that the gauge transformations
satisfy an additional condition
$\llp(f\ot f\sp{-1})\ci\de\lrp (M_{A})\inc N_{P}\,$.
(A related discussion can be found around formula (48)
in \cite{bmq}; note that this condition
is satisfied in the classical situation,
in which $M_{A}=(\mbox{Ker}\,\epsilon )\sp{2}$ ---
see Example 1 on p.132 in \cite{wor}.) Therefore,
 when dealing with quantum gauge transformations, we will assume the universal
differential calculus.

\bpr\label{gact}
Let $f\in GT(P)$ and $\omega\in\cal C(P)$, where $\cal C(P)$ denotes the space of
all connection forms on a quantum principal bundle $(P,A,\dr ,0,0)$. 
Then the formula
\[
G\!\sb f\omega =f*\omega * f\sp{-1} + f*(d\ci f\sp{-1})
\]
(cf.~(20) in \cite{wz}) defines an action $G:GT(P)\times\cal  C(P)\lra\cal C(P)\,$.
\epr
{\it Proof.} Let us verify first that $G\!\sb f\omega$ is indeed a connection form.
\be
\item $G\!\sb f\omega (k) = 0$. Obvious.
\item Taking into account that \T\ is a
left $P$-module morphism (see Point~4 in
Definition~\ref{qpbdef}) and remembering that for the universal differential
calculus $N\sb P\! =\! 0\! =\! M\sb A$, one can see that

\bea
&& (\T\ci G\sb{\! f}\omega)(a) \\
&& =  f(a\sb{(1)})(\T\cc\omega)(a\2)
(\Delta\sb R\cc f\sp{-1})(a\3 ) \\
&& + \llp(m\ot id)\ci (id\ot\dr)\lrp\llp
f(a\sb{(1)})\ot
 f\sp{-1}(a\2) - \epsilon(a)\ot 1\lrp \\
&& = \lblp f(a\sb{(1)})\ot
\llp a\2-\epsilon(a\2)
\lrp\lbrp\llp f\sp{-1}(a\4 )\ot
S(a\3 )a\sb{(5)}\lrp \\
&& + f(a\sb{(1)})f\sp{-1}(a\3 )\ot
S(a\2)a\4  - 1\ot\epsilon(a) \\
&& = f(a\sb{(1)}) f\sp{-1}(a\3 )
 \ot\llp\epsilon(a\sb {(2)})
a\4  - S(a\2)a\4 \lrp \\
&& + f(a\sb{(1)})f\sp{-1}(a\3 )\ot
S(a\2)a\sb
{(4)}  - 1\ot\epsilon(a) \\
&& = 1\ot\llp a-\epsilon(a)\lrp .
\eea
\item $\dsr\ci G\sb{\! f}\omega = (G\sb{\! f}\omega\ot id)\ci ad\sb R$
can be proved much as Proposition~\ref{gtprop}.
\ee
Now, to complete the proof, it suffices to note that
$G\sb{\! f*g}\omega = G\sb{\! f}G\sb{\! g}\omega$.
\hsp{\fill}$\Box$
\bigskip

The action of the gauge group $GT(P)$ on the space of connections can be derived
from its action on connection forms. It is explicitly described by

\bpr\label{gtpi}
Let $GT(P)$ be as in Proposition~\ref{gact}.
Denote by ${\cal P}(P)$ the space of
all connections on $P$, and by $\Upsilon$ the bijection providing the
correspondence between connections and connection forms, i.e., let
\[
\Upsilon : {\cal P}(P)\ni\Pi\longmapsto\sigma\sb\Pi\ci\llp 1\ot
 (id -\epsilon )\lrp\in\cal C(P)\, ,
\]
where $\sigma\sb\Pi :
P\ot\mbox{\em Ker}\,\epsilon\ra\Omega\sp 1\! P$ is
the unique left \mbox{$P$-module} homomorphism satisfying
 \mbox{$\mbox{\em\sf T}\ci\sigma\sb\Pi =id$}  and
\mbox{$\sigma\sb\Pi\ci\mbox{\em\sf T}=\Pi$}.
{\em\footnote{ 
See Proposition~4.4 and the paragraph above it in \bmq .}}
Then the map
${\cal G} : GT(P)\times{\cal P}(P)\lra{\cal P}(P)$ given by
\beq\label{gpi}
{\cal G}\sb f\Pi
= m\sb{\Omega\sp 1P}\ci\lblp id\,\ot\,\llp\Pi\cc d\cc
(id\! *\sb R\! f)
\lrp\! *\sb R\! f\sp{-1}\lbrp +
 m\sb{\Omega\sp 1P}\ci\llp id\,\ot\, f\! *\! (d\!
\ci\! f\sp{-1})\lrp\ci\mbox{\em\sf T}
\eeq
is the action of the gauge group $GT(P)$ on the space of connections
${\cal P}(P)$, and the following diagram commutes:
\bcd
GT(P)\times{\cal P}(P) @>(id,\Upsilon)>> GT(P)\times\cal C(P)\\
@V{\cal G}VV @VGVV\\
{\cal P}(P) @>\Upsilon >>\cal C(P)\\
\ecd
\epr
{\it Proof.} Both assertions of the proposition follow from the formula:
\beq\label{gtpifor}
\mbox{\large$\forall$}\,
f\in GT(P),\,\Pi\in{\cal P}(P):\; {\cal G}\sb f\Pi =
(\Upsilon\sp{-1}\ci G\!\sb f\ci\Upsilon) (\Pi )
\eeq
Since both sides of the above equation are left \mbox{$P$-module}
homomorphisms, in order to prove~(\ref{gtpifor}),
it suffices to show that,
for any $f\in GT(P)$ and $\Pi\in\cal P(P)$, we have
\[
(\cal G\sb f\Pi )\ci d =
\llp(\Upsilon\sp{-1}\cc G\!\sb f\cc\Upsilon) (\Pi )\lrp\ci d\, ,
\]
that is, 
\bea\label{gtpifor2}
\hsp{2.25mm} &&
 m\sb{\Omega\sp 1P}\ci\Lblp id\,\ot\, f\! *\!\Llp\sigma\sb\Pi\cc
\llp 1\te (id - \epsilon )\lrp\Lrp\! *\! f\sp{-1}\Lbrp\ci\mbox{\sf T}
\ci d\hsp{57mm}
\mbox{(\theequation )}\addtocounter{equation}{1} \\
&&
 =  m\sb{\Omega\sp 1P}\ci\lblp id\,\ot\,\llp\Pi\cc d\cc
(id\! *\sb R\! f)
\lrp\! *\sb R\! f\sp{-1}\lbrp\ci d
\hsp{29.5mm} \mbox{(pure gauge terms cancel).}
\eea
To do so, let us calculate the value of the left hand side of
 (\ref{gtpifor2})
on arbitrary $p\in P$:
\bea
&&
\LAblp m\sb{\Omega\sp 1P}\ci\LAlp id\,\ot\, f\! *\!\Llp\sigma\sb\Pi\cc
\llp 1\te (id - \epsilon )\lrp\Lrp\! *\! f\sp{-1}\LArp\ci\mbox{\sf T}
\ci d\LAbrp (p)
\\ &&
=\LAblp m\sb{\Omega\sp 1P}\ci\LAlp id\,\ot\, f\! *\!\Llp\sigma\sb\Pi\cc
\llp 1\te (id - \epsilon )\lrp\Lrp\! *\! f\sp{-1}\LArp\LAbrp
(p\0\ot p\sb{(1)} - p\ot 1)
\hsp{11mm} \mbox{(see (24)\bmq )}
\\ &&
=p\0f(p\sb{(1)})\sigma\sb\Pi\llp1\ot p\2-\epsilon (p\2)
\ot 1\lrp f\sp{-1}(p\3 )
\\ &&
=\sigma\sb\Pi\llp p\0f(p\sb{(1)})\ot p\2-
p\0f(p\sb{(1)})\epsilon (p\2)
\ot 1\lrp f\sp{-1}(p\3 )
\\ &&
=\llp\sigma\sb\Pi\ci\mbox{\sf T}\ci d\ci
 (id\! *\sb R\! f)\lrp (p\0)\,
f\sp{-1}(p\sb{(1)})
\hsp{41 mm} (\mbox{$\, id\! *\sb{\! R}\! f$ is right-covariant})
\\ &&
=\lblp\llp\Pi\cc d\cc
 (id\! *\sb R\! f)\lrp*\sb Rf\sp{-1}\lbrp\, (p)
\\ &&
 = \Lblp m\sb{\Omega\sp 1P}\ci\Llp id\,\ot\,\llp\Pi\cc d\cc
(id\! *\sb R\! f)
\lrp\! *\sb R\! f\sp{-1}\Lrp\Lbrp (1\ot p - p\ot 1)
\\ &&
 = \Lblp m\sb{\Omega\sp 1P}\ci\Llp id\,\ot\,\llp\Pi\cc d\cc
(id\! *\sb R\! f)
\lrp\! *\sb R\! f\sp{-1}\Lrp\ci d\Lbrp (p)
\eea

Hence, we can conclude that (\ref{gtpifor2}) is true,
and the proposition
follows. \hfill{$\Box$}

\bre{\em
The left \mbox{$B$-module} isomorphism $(id*\sb Rf):P\lra P$
can be regarded as
a quantum version of a gauge transformation 
 understood as an appropriate
diffeomorphism of the total space of a principal bundle
(see p.339 in \cite{booss} and Definition~5.1 in~\cite{tbt}).
In fact, the map $f\to id*\sb Rf$ is a group isomorphism
(see Corollary~5.3 in \cite{tbt}).
(Observe that $(id*\sb Rf\sp{-1})\ci
(id*\sb Rf)=id=(id*\sb Rf)\ci (id*\sb Rf\sp{-1}).\,$)
\hfill{$\Diamond$}}\ere

Having defined and described quantum gauge
transformations and their
action on the space of connections and connection forms,
 we can now show that these
transformations preserve the strongness of
 a connection, i.e.
(provided $M\sb A= 0 =N\sb P$) their action
is well-defined on the space
\spp\ of all strong connections  on a
quantum bundle $P$. (Obviously, their action
 is then also well-defined on the space $\cal S\cal C(P)$
of all strong connection forms.)

\bpr
Let $P(B,A)$ be a quantum principal
 bundle with the universal differential calculus. Then
$\mbox{\large$\forall$}\, f\in GT(P)$:
 $\Pi\in\spp \,\Leftrightarrow\, \cal G\sb f\Pi \in\spp$.
\epr
{\it Proof.}
Note first that, for any right $A$-comodule coaction
\mbox{$\rho\sb{\cal C}:{\cal C}\lra{\cal C}\ot A$}, any \linebreak
\mbox{$\alpha\sb 1\in\mbox{Hom}\sb k\llp {\cal C},\Omega
\sp m(X)\lrp$}
and \mbox{$\alpha\sb 2\in\mbox{Hom}\sb k\llp{\cal C},
\Omega\sp n(X)\lrp$} (where $m\geq 0,\, n\geq 0$
and $\Omega (X)$ is any differential algebra), there is the familiar
graded Leibniz rule:
\[
d\ci (\alpha\sb 1\! *\sb{\rho\sb{\cal C}}\!\alpha\sb 2)=(d\ci\alpha\sb
 1)*\sb{\rho\sb{\cal C}}\alpha\sb 2
+ (-)\sp m\alpha\sb 1*\sb{\rho\sb{\cal C}}(d\ci\alpha\sb 2)
\]
Now, using similar calculations as in the proof of Proposition~\ref{gtpi} 
and the Leibniz rule for the convolution, one can see that,
for any $f\!\in\!
 GT(P),\,\Pi\!\in\! {\cal P}(P)$,
\bea
\hsp{20mm} && (id-{\cal G}\sb f\Pi)\ci d
\hsp{90mm} \mbox{(see (\ref{gpi}))}
\\ &&
= d - \llp\Pi\ci d\ci (id\! *\sb R\! f)\lrp *\sb Rf\sp{-1}
- id*\sb Rf*(d\cc f\sp{-1}) \\
\hsp{14mm} && = \llp d\ci (id\! *\sb R\! f)\lrp\, *\sb R\, f\sp{-1}
- \llp\Pi\ci d\ci (id\! *\sb R\! f)\lrp *\sb Rf\sp{-1}\\
\hsp{14mm} && = \llp (id-\Pi )\ci d\ci
(id\!*\sb R\! f)\lrp\, *\sb R\, f\sp{-1} \, .
\eea
Hence, $\mbox{\large$\forall$}\,
f\!\in\! GT(P):\,\Pi\!\in\!\spp \Rightarrow {\cal G}\sb
f\Pi\!\in\!\spp $, and since ${\cal G}$ is a
group action on ${\cal P}(P)$, $\mbox{\large$\forall$}
f\!\in\! GT(P):\, {\cal G}\sb f\Pi\!\in\!\spp
\Rightarrow \Pi\!
\in\!\spp $. \hfill $\Box$
\bigskip

\section{Curvature}

We are ready now to examine the properties of the exterior
 covariant derivative (see Definition~\ref{Ddef})
associated with a strong connection.
 This will lead to the definition of a (global)
 curvature form on a quantum principal bundle. 
The following proposition and corollaries describe
the composition of the covariant exterior derivative with strongly
tensorial differential forms
(see Definition~\ref{tfdef}). They and Proposition~\ref{curadj} are analogous
to the corresponding local (i.e.~valid only for trivial quantum bundles)
statements made in \bmq . 
As we do not know how to characterize all
 differential algebras for which $D$ can be well-defined,
we will simply assume here, whenever needed, that $\Omega(P)$ 
is a right-covariant differential algebra such that $D$ is well-defined. 

\bpr  [cf.\ (17) and (76) in \bmq ]\label{Dformula}
Let $(P,A,\dr ,N\sb P,M\sb A)$ be a quantum principal bundle with a
connection form $\omega$, and let $\hO (P)$ be a differential algebra
such that\linebreak 
$\op = \hO\sp1\! P/N\sb P$ and the exterior covariant derivative
$D\sp\omega$ associated to \ho\ is well-defined by formula~(\ref{689}). 
Then, for all  
\mbox{$\phi\in ST\sb\rho\llp V,\Omega\sp n(\! P)\lrp ,\, n\in \{0\}\cup{\Bbb N}$},
\[
 D\sp\omega\ci\phi =d\ci\phi - (-)\sp n \phi *\sb\rho\omega\, .
\]
\epr
{\it Proof.}
Note that, for any $d\alpha\in\Omega\sp n(B)$,
$n\in{\Bbb N}$, $p\in P$, we have
\bea
\hsp{6mm}&& (d-D\sp\omega)(d\alpha .p)
\\ &&
=(-)\sp nd\alpha dp - (-)\sp nd\alpha .(id - \Pi\sp\omega )(dp)
\hsp{70mm}\mbox{(see (\ref{confor}))}
\\ &&
=(-)\sp nd\alpha .p\0\omega(p\sb{(1)})
\\ &&
=(-)\sp n(id*\sb{\cal R}\omega )(d\alpha .p)\, .
\eea
On the other hand, for all $v\in V$, $\,\phi (v)$
can be written as a finite sum
$\sum\sb id\alpha\sb i.p\sb i$ for some closed differential forms
$d\alpha\sb i\in\Omega\sp n(B)$ and 0-forms
$p\sb i\in P$. Hence, with the help of Lemma~\ref{rcovlem}, we obtain
\[
 \llp(d-D\sp\omega)\cc\phi\lrp\, (v)
=\llp(-)\sp n(id*\sb{\cal R}\omega )\cc\phi\lrp\, (v)
=(-)\sp n(\phi *\sb\rho\omega )(v)\, ,
\]
and the assertion follows.
\hfill{$\Box$}

\bco [cf.\ (7)\cite{bmq}]\label{Dcor}
Let $\hO (P)$ be as in the proposition above. 
If $D$ is the exterior covariant derivative
associated to a strong connection, then, for every \nin ,
we have $D\ci\lblp ST\sb\rho\llp V, \Omega\sp n(\! P)\lrp\lbrp\inc ST\sb\rho\llp
V,\Omega\sp{n+1}(\! P)\lrp .$
\eco
{\it Proof.}
In the same way as in the case of the differential envelope, it can be directly
calculated that $D$ is always right-covariant (see Appendix~A in \bmq ), and thus,
for all connections, $D$ composed with a pseudotensorial differential
 form is a tensorial differential form, i.e. 
\[
\mbox{\large\boldmath$\forall$}\, n\in\{ 0\}\!\cup\!{\Bbb N} :\;
D\ci\lblp PT\sb\rho\llp V, \Omega\sp n(\! P)\lrp\lbrp\inc T\sb\rho\llp
V,\Omega\sp{n+1}(\! P)\lrp\, .
\]
 Hence, to prove the assertion of this corollary,
it suffices to note that, if $D$ is associated to a strong connection, then,
as can be seen from the second line of the first calculation in
Proposition~\ref{Dformula}, $D$ evaluated on a strongly horizontal differential
form yields a strongly horizontal differential form.
\hfill{\mbox{$\Box$}}

\bre\label{scdefrem}\em
With the help of $D$ one can equivalently define a strong
connection as a connection
whose exterior covariant derivative maps $P$ into strongly horizontal forms,
i.e.\ $D(P)\inc\Omega\sp 1\sb{shor}(P)$. Obviously, since $D$ can be defined on $P$
for any differential calculus, such a definition works on any quantum
principal bundle.
\hfill{$\Diamond$}\ere

\bco [cf.\ Section~3 in \cite{bmq}]\label{D2cor}
Let \ho\ be a strong connection form, and let $D\sp\omega$ and $\hO (P)$ be as in 
Proposition~\ref{Dformula}. Then
\beq\label{D2gen}
\mbox{\large$\forall$}\,\phi\in ST\sb\rho\llp
V,\Omega(P)\lrp\; :\; (D\sp\omega)\sp 2\ci\phi
= -\phi *\sb\rho(d\ci\omega +\omega*\omega).
\eeq
\eco

Formula (\ref{D2gen}) is, in particular, true for a classical principal bundle
$\widetilde{P}(M,G)$. (A classical principal bundle is a special case of a
quantum principal bundle when we replace the algebraic tensor product by the
appropriately completed tensor product.)
It is a generalization of the classical formula
\beq\label{D2clas}
D\sp 2\alpha =
\varrho\sp{'}(F)\wedge\alpha\, ,
\eeq
 where $\alpha$ is a differential form on $\widetilde{P}$
 with values in a finite dimensional vector space $W$, the homomorphism
\mbox{$\varrho\sp{'} : \frak g\ra\frak g\frak l(W)$} 
is the Lie algebra representation induced
by a homomorphism \mbox{$\varrho :G \ra GL(W)$},
and $F$ is the curvature form of a connection defining $D$ (e.g., 
see (19) in~\cite{tr}). More precisely,
 taking $V$ to be the tensor algebra of the dual vector space
$W\sp*$, defining $\rho\sb R$~by
\[
\mbox{\large$\forall\,$} v\in W\sp*\, :\,
\rho\sb R(v)(w,g)=v\llp\varrho(g\sp{-1})w\lrp\, ,
\]
putting 

\[
\phi\sb\alpha (v) : 
\left\{\begin{array}{ll}
{\displaystyle\bigwedge\sp{\deg\alpha}T\sb
{\tilde{p}}\widetilde{P}\ni X\sb{{\tilde{p}}}
\longmapsto\left\{
\begin{array}{ll}
(v\cc\alpha) (X\sb{{\tilde{p}}})
& \mbox{for $v\in W\sp*$}\\
0 & \mbox{otherwise}
\end{array}
\right.} & \mbox{for $\deg\alpha > 0$}\\ 
\phantom{.} & \phantom{.}\\
{\displaystyle\widetilde{P}\ni\widetilde{p}
\longmapsto(\check{v}\cc\alpha)
(\widetilde{p})\in\Bbb R}
 & \mbox{for $\deg\alpha = 0\,$,}\\
\end{array}\right.
\]\ \\

where $\check{v}$ is the polynomial function on $W$ corresponding to the tensor $v$,
and remembering that every $\frak g$-valued
differential form $\vartheta$ on $\widetilde{P}$ can be viewed as an
\mbox{$\epsilon_{C\sp{\!\infty\!}(\! G\!)}$-d}erivation
\mbox{$\psi_{\vartheta} : C\sp{\infty}(G)\ra\bigwedge\sp*(\widetilde{P})$} 
given by $\psi_{\vartheta}(a)X_{\tilde{p}} = \vartheta(X_{\tilde{p}}) a$
(or $\psi_{\vartheta}(a)\widetilde{p} =\vartheta (\widetilde{p}) a$
 if $\deg\alpha = 0$), we can rewrite (\ref{D2clas}) as
\[
D\sp 2\cc\phi\sb\alpha =
-\phi\sb\alpha\! *\sb\rho\! F\, .
\]
Note that, since every vector space
$W$ is a Lie group and there is a canonical
isomorphism between $W$ and the Lie algebra of $W$,
we could define $\phi\sb\alpha$ in the
same way as we define~$\psi\sb\vartheta$. But then we would not have
$\phi\sb\alpha (1) = \delta\sb{0,\deg\alpha}$, which we need in Appendix~A
(see Proposition~\ref{asssecprop}).
In the classical case, we can replace $F$ by $\; d\ci\omega +
\omega*\omega\;$ or
$\; D\ci\omega\;$ in the last formula, but we need to put
$\; F=d\ci\omega + \omega *\omega\;$  to obtain (\ref{D2gen}).
Also, formula (\ref{D2clas}) can be considered as a motivating factor in defining
 the curvature of a connection on a projective module 
as the square of a covariant derivative
(see p.554 in \cite{conbook}; 
covariant derivative $\equiv$ connection on a projective
module). Therefore, it is natural to define the (global)
curvature form of a connection $\Pi\sp\omega$
in the following way:

\bde \label{curdef}
Let $\omega$ be a connection \mbox{1-form} on a
quantum principal bundle. The
differential form \mbox{$d\cc\omega +
\omega\! *\!\omega$} is called the
curvature form of $\omega$ and is denoted by
$F\sb\omega\,$.
\ede
 Clearly, if $\omega$ is a strong connection
 form, then, at least in the case of a trivial
 bundle, for any differential algebra
$\Omega(P)$, even if $D$ is not well-defined,
 $F\sb\omega$ is horizontal
(\mbox{$F\sb\omega =\Phi\sp{-1}\!\! *\! F\sb
 \beta\! *\!\Phi$},
where $\Phi$ and $\beta$ are as in
Proposition~\ref{scprop} and $F\sb\beta :=
d\cc\beta +\beta\! *\!\beta$
is a local curvature
form, cf.~(11)\cite{bmq}). Moreover, unlike the
expression $D\cc\omega$,
the so-defined
curvature form has the desired (at least from
the point
 of view of Yang--Mills theory) transformation
properties, i.e.~we have:
\nopagebreak
\bpr  [cf.\ (13)\cite{bmq}, (20)\cite{wz}]\label{curadj}
Let $P(B,A)$ be a quantum principal bundle with
the universal differential calculus,
$\omega\in\cal C(P)$ and $f\in GT(P)$. Then
\[
F\sb{G\!\sb f\omega} =
f*F\sb\omega*f\sp{-1}\, .
\]
\epr
{\it Proof.} Straightforward.\hfill{$\Box$}

\section{\mbox{\boldmath$U\sb q(2)$}--Yang--Mills Theory on a Free Module}

To begin with, let us show that it is possible to define an action functional
on quantum bundle connections in such a way that, at least in the case of
a trivial quantum bundle, 
it agrees with the Yang--Mills
 action functional constructed in Section~1 of both \cite{conri} and \cite{ri}.
(Clearly, we assume that the `base space' of a quantum bundle is an algebra over
 which modules are considered in~\cite{conri,ri}.)
Considerations in \cite{bmc} and Proposition~\ref{curadj}
 suggest that, if $A$ is a matrix quantum group and $T$ its matrix of generators
(fundamental representation, cf.\ p.628 in~\cite{cmp}),
it is reasonable to define 
an action on a quantum principal bundle $P(B,A)$ to be
(compare with the Lagrangian given by (6.65) in \cite{md2} or (4.1) in \cite{md4})
\beq\label{ym}
YM(\omega) = - \llp{\cal T}\ci Tr\ci (F\sb\omega*F\sb\omega)\lrp(T),
\eeq
where 
$Tr$ is the usual matrix trace, and
\mbox{${\cal T}:\Omega P\ra k$} is a linear map vanishing on
$[P,\Omega P]$. (To ensure that we have an ample supply
of connections, throughout this section we will
use the universal differential calculus.)
Clearly, remembering the property of $T$ that,
for any $\mbox{f}\sb 1$ and $\mbox{f}\sb 2\,$,
$
(\mbox{f}\sb 1\! *\! \mbox{f}\sb 2)(T\sb{ab})=\sum\sb c
\mbox{f}\sb 1(T\sb{ac})\, \mbox{f}\sb 2(T\sb{cb})\, ,
$
one can see that
\[
\mbox{\large$\forall$}\,
f\!\in\! GT(P) , \omega\!\in\!\cal C(P) :\,
YM(G\!\sb f\omega) = YM (\omega)\, .
\]
Similarly, if $P(B,A)$
is a trivial bundle with a trivialization
$\Phi$ (which we will also assume for
the rest of this section), then
for any $\omega\,$,
\beq\label{ymbeta}
YM(\omega) = -\llp {\cal T}\ci
Tr\ci(\Phi*F\sb\omega*F\sb\omega
*\Phi\sp{-1})\lrp(T) =
-\llp {\cal T}\ci Tr\ci
(F\sb\beta*F\sb\beta)\lrp(T) =:YM(\beta )\, ,
\eeq
 where $\beta$ is given by formula (\ref{betadef})
(see Remark~\ref{secrem}) and $F\sb\beta $
is the curvature form associated to it (see the end of the previous section).
As to \ct ,  observe that one should expect to have a lot of information
vested in it: projection from the universal to a non-universal calculus and
metric (Hodge star). In what follows, we specify \ct\ in such a way as to
incorporate the Yang--Mills functional presented in \cite{conri,ri} into the quantum
bundle framework. One ought to bear in mind that to obtain a $q$-deformed
Yang--Mills theory in the spirit of quantum groups, one should invent another
\ct\ or change the formula (\ref{ym}) altogether. Here, however, we investigate
what effects can entail from the deformation of the structure group alone.

\bde\label{tld}
Let \ctb\ be a faithful invariant trace on $B$ (as in \cite{conri,ri}),
$\cal L$ be a finite dimensional Lie subalgebra of $\,\mbox{\em Der}(B)$, 
$\{ X\sb l\}\sb{l\in\{1,...,\mbox{\em\scriptsize dim}\cal L\}}$ be its basis, and
$\widetilde{X}\sb l=X\sb l\ot id$,\linebreak
 $l\in\{1,...\, ,\mbox{\em dim}\,\cal L\}$. 
We put
\[
\ct\ (\alpha )=
\left\{
\begin{array}{ll}
\hsp{3mm}(\ctb\te\epsilon)\; (\alpha) & \mbox{if $\deg\alpha =0$}\\
\phantom{1} & \phantom{1}\\
\hsp{3mm}\displaystyle 0 & \mbox{if $\deg\alpha =2m-1$ }\\
\phantom{1} & \phantom{1}\\
{\displaystyle\sum\sb{l\sb 1<\cdots <l\sb m}}
\llp(\ctb\te\epsilon )\ci\alpha\lrp\,
(\widetilde{X}\sb{l\sb 1}\wedge\!\cdots\!\wedge\! \widetilde{X}
\sb{l\sb m}\te
\widetilde{X}\sb{l\sb 1}\wedge\!\cdots\!\wedge\!
\widetilde{X}\sb{l\sb m}) &
\mbox{if $\deg\alpha = 2m$}\, ,
\end{array}\right.
\]
\smallskip
where $m\!\in\!{\Bbb N}$.
\ede

Recall that the product of differential \mbox{1-forms}
evaluated on the tensor product of
derivations is, as in the classical case, given by
\[
(\alpha\sb 1\!\cdots\!\alpha\sb n)\, (Y\sb 1\ot\!\cdots\!\ot Y\sb n)=
\alpha\sb 1(Y\sb 1)\cdots\alpha\sb n(Y\sb n)
\]
and $(b\sb 0db\sb 1)(Y)=b\sb 0Y(b\sb 1)$ (see p.403 in \cite{dv}).
(Observe that, as $\{ Y\sb i\}
\sb{i\in\{ 1,\cdots ,n\}}$ are derivations,
it does not matter in which way
we write a differential form as a sum of products of differential
 \mbox{1-forms}.)
Now, with \ct\ defined as above, we have $\ct\ ([P,\Omega P])=0$. 
Moreover, we have

\bpr\label{ymprop}
Let $A$ and $T$ be as above. Also, let $B$ be a $*$-algebra, 
$P(B,A)$ a trivial quantum principal bundle with a 
trivialization~$\Phi$ (e.g., $P=B\te A$), \ho\ a strong
connection form on $P$, and 
\ct\  as in Definition~\ref{tld}. Then
\beq\label{ymeq}
YM(\ho)=YM(\nabla\sp\omega)\, ,
\eeq
where $\nabla\sp\omega=d+\hb (T)$ is the \ho\ induced connection
on the right module $B\sp n$ (\hb\ is given by~(\ref{betadef}), $n$~is 
the size of~$T$, $\nabla\sp\omega\xi=d\xi +\hb (T)\xi$ --- see p.637 in \bmq ), 
and the right hand side of (\ref{ymeq}) is defined
in Definition~\ref{ymfdef}.
\epr

{\it Proof.}
Note first that, by Proposition~\ref{scprop}, 
$\hb (T)\in M\sb n(\hO\sp1\! B)$. Consequently, 
$F\sb\beta (T)= \linebreak 
d\beta (T)+\beta (T)\sp 2$ is an element of
$M\sb n(\hO\sp2\! B)$ (see the paragraph above
Proposition~\ref{curadj} for $F\sb\beta$).
Therefore, since the curvature $(\nabla\sp\omega)\sp 2$
equals just \mbox{$d\beta (T)+\beta (T)\sp 2$} (cf.~p.681 in \cite{conact}),
with the help of (\ref{ymbeta}), Corollary~\ref{traceco} and 
Proposition~\ref{spera}, we have
\bea
\hsp{10mm}YM(\ho )&=&YM(\beta ) 
\\ &=&
-\sum\sb{r<s}\llp (\cal T\sb B\te\epsilon )\ci 
Tr(F\sb\beta *F\sb\beta)(T)\lrp (\widetilde{X}\sb
r\wedge\widetilde{X}\sb s\ot\widetilde{X}\sb
r\wedge\widetilde{X}\sb s)
\\ &=&
-\sum
\sb{{\scriptstyle i,j}\atop
{\scriptstyle r<s}}
\lblp\ctb\cc\llp F\sb\beta (T\sb{ij})
F\sb\beta (T\sb{ji})\lrp\lbrp
(X\sb r\!\wedge\!
X\sb s\te X\sb r\!\wedge\!
X\sb s)
\\ &=&
-(\ctb\cc Tr)\,\llp\sum\sb{r<s}F\sb\beta (T)(X\sb r\wedge X\sb s)
F\sb\beta (T)(X\sb r\wedge X\sb s)\lrp
\\ &=&
-\cal T\sb E\Llp\sum\sb{r<s}\llp (\nabla\sp\omega)\sp2(X\sb r\wedge
X\sb s)\lrp\sp2\Lrp
\\ &=&
-\cal T\sb E\{\Theta\sb{\nabla\sp\omega}, \Theta\sb{\nabla\sp\omega}\}
= YM(\nabla\sp\omega). \hsp{73mm}\Box
\eea\ 

Thus, the Yang--Mills
action functional given by (\ref{ym}) coincides, as stated above,
 with the Yang--Mills action
functional defined in \cite{conri,ri} (cf.\ Section~VII.D in
\cite{dvkmm} and Section~V.B in~\cite{dkm}).
 This confirms that, as was mentioned in
\bmq , the formalism of quantum group gauge theory is ``not incompatible with the
existing ideas in non-commutative geometry".

\bre\label{trem} \em
Observe that although any trivial quantum bundle $P(B,A)$
is isomorphic with
$B\ot A$ (algebras $B$ and $B\ot 1$ identified) as a left
\mbox{$B$-module}, we cannot claim in general that it is
isomorphic with
$B\ot A$ as an algebra (see the first paragraph of Section~3.1 in
\cite{mp}). But here, since our attention is restricted to
strong connections, the structure of the `total space'
is not important
--- we can equally well define the Yang--Mills action functional by
\[
YM(\beta) = -\llp {\cal T}\sb{\Omega B}\ci  Tr\ci
(F\sb\beta*F\sb\beta)\lrp(T)\, ,
\]
where ${\cal T}\sb{\Omega B}$ is a \mbox{$k$-linear} map vanishing on
$[B,\Omega B]$. (Obviously, the Yang--Mills action thus defined
is invariant under
the local gauge transformations, i.e.~gauge transformations
taking values in $B$
rather than in $P$; see Definition~\ref{gtrans}).
 \hfill{$\Diamond$}\ere

\bre\label{chern} \em
Note that the derivations used in Definition~\ref{tld}
are not fully antisymmetrized --- there is a tensor product
in between two groups of antisymmetrized derivations. This is
a key difference between the construction of $YM(\nabla )$ and
the construction of the second Chern class. (One should think of
$\cal T$ defined in Definition~\ref{tld} as a pairing rather than
a trace.) 
 \hfill{$\Diamond$}\ere

Now, we are going to take a closer look at what happens
 if we choose $A=U\sb q(2)$  (with $q\in {\Bbb R}_{+}$).
(We continue to assume, as in Proposition~\ref{ymprop}, that B is a $*$-algebra,
so that $\hO (B)$ is also a $*$-algebra; cf.~(1.32) in~\cite{wor}.) 
To do so, let us first recall the definition of $U\sb q(2)$
(see Lecture~5 in \cite{ta}) :

\bde\label{uqdef}
$(U\sb q(2),\Delta ,\epsilon ,S)$ is a matrix
$*$-Hopf algebra determined by the following equalities:
\[ U\sb q(2) = {\Bbb C}\langle a,b,t,a\sp*,
b\sp*,t\sp*,1\rangle /({\cal J}+{\cal J}\sp*),\]
where
\bea
{\cal J} & =  &(ab - qba, bb\sp* -b\sp*b, ab\sp*
 -qb\sp* a, a\sp*a-aa\sp* +(q\sp{-2}-1)bb\sp*,\\
& & ta-at,tb-bt,ta
\sp*-a\sp*t,tb\sp*-b\sp*t,tt\sp*-t\sp*t,
 aa\sp* +bb\sp* -1,tt\sp* -1)
\eea
and $q\in {\Bbb R}\smallsetminus\{0\};$ {\em\footnote{ Since,
further on, we want $\mbox{\scriptsize $\pmatrix{1&0\cr 0&q}$}$ to be
positive definite, we actually need to assume
that $q>0$.}}
\[\Delta T =T\buildrel .\over\ot  T,\;\;
\epsilon (T) =I,\;\; S(T)=T\sp\dagger ,\]
where
\[
T=\pmatrix{a&b\cr -q\sp{-1}b\sp*t\sp* &
 a\sp*t\sp*}\,
 \]
and $\stackrel{.}{\ot}$ is the  matrix
multiplication with the product of its
entries replaced by the tensor product $\ot\,$;
and
\[
\Delta t =t\ot t , \;\; \epsilon(t) =1
,\;\; S(t)=t\sp*,\;\; S(t\sp*) =t.
\]
\ede

The next step is to establish an equivalence between a certain
class of strong connections on a trivial $U\sb q(2)$-bundle
and the space of hermitian connections on the associated free module.
Since any $\beta\in
\mbox{Hom}\sb{{\Bbb C}}(U\sb q(2),\Omega B)$ satisfying
$\beta({\Bbb C})=0$ is
a connection form on the base space $B$ of a
 trivial $U\sb q(2)$ bundle (see Remark~\ref{secrem}
 and the paragraph below) and the
$\beta$-induced connection $\nabla$ on $B\sp 2$
 depends only on $\beta(T)$, we will
restrict our attention to connection forms
that are, in some way, uniquely
determined by $\beta(T)$. In the classical situation, $\beta$ is an
\mbox{$\epsilon$-derivation} and thus is automatically
determined by $\beta (T)$.
Mimicking this classical differential geometry formula for $\beta$,
we define an auxiliary map 
\mbox{$
\widetilde\hb :\Bbb C\langle a,b,t,a\sp*, b\sp*,t\sp*,1\rangle
\ra\hO\sp1\! B
$}
by\\

\begin{eqnarray}\label{betafor}
&& 
\hsp{-2mm}\widetilde{\beta}(
a\sp{k\sb1}a\sp{*\, l\sb1}b\sp{m\sb1}b\sp{*\, n\sb1}t\sp{p\sb1}t\sp{*\, r\sb1}
\cdots 
a\sp{k\sb s}a\sp{*\, l\sb s}b\sp{m\sb s}b\sp{*\, n\sb s}t\sp{p\sb s}t\sp{*\, r\sb s})
\nonumber \\ && \ \nonumber \\ &&= 
\left\{\begin{array}{ll}
0 &\mbox{for $\displaystyle\sum\sb{i=1}\sp sm\sb i+n\sb i\geq 2$ }
\\ \ & \ \\
q\sp{\sum\sb{i=j+1}\sp s(l\sb i-k\sb i)}\;\widetilde{b} 
& \mbox{for $m\sb j=1$, $j\in\{ 1,...,s\}$, 
$\displaystyle\sum\sb{i=1}\sp sm\sb i+n\sb i=0$\phantom{........}}
\\ \ & \ \\
q\sp{\sum\sb{i=j+1}\sp s(l\sb i-k\sb i)}\;\widetilde{b}\,\sp* 
&\mbox{for $n\sb j=1$, $j\in\{ 1,...,s\}$, 
$\displaystyle\sum\sb{i=1}\sp sm\sb i+n\sb i=0$ }
\\ \ & \ \\
{\displaystyle\sum\sb{i=1}\sp sk\sb i\widetilde{a}
+\sum\sb{i=1}\sp sl\sb i\widetilde{a}\,\sp*
+\sum\sb{i=1}\sp sp\sb i\widetilde{t} 
+\sum\sb{i=1}\sp sr\sb i\widetilde{t}\,\sp*}
& \mbox{for $\displaystyle\sum\sb{i=1}\sp sm\sb i+n\sb i=0\, ,$}
\end{array} \right .
\end{eqnarray}\ \\

where, a priori, $\widetilde{a}\, ,\,\widetilde{b}$ and $\widetilde{t}$
are any differential forms in $\Omega\sp 1\! B$. 
One can verify that\linebreak
 \mbox{$\widetilde\hb (\cal J+\cal J\sp*)=0$} if
$\widetilde{a}+\widetilde{a}\,\sp*=0=
\widetilde{t}+\widetilde{t}\,\sp*$. Hence, since
\[
\widetilde\beta 
(aa\sp*+bb\sp*-1)=0=\widetilde\beta (tt\sp*-1)\;\;\mbox{\boldmath$\Rightarrow$}
\;\;\widetilde{a}+\widetilde{a}\,\sp*=0=\widetilde{t}+\widetilde{t}\,\sp*\, ,
\]
we can conclude that (\ref{betafor})
defines $\hb\in\mbox{Hom}\sb{\Bbb C}(U\sb q(2),\hO\sp1\! B)$ such that
$\hb (1)=0$
(i.e., a \mbox{$U\sb q(2)$-connection} form) if and only if
$\widetilde{a}+\widetilde{a}\,\sp*=0=\widetilde{t}+\widetilde{t}\,\sp*$.
Moreover, it is straightforward to check that
\beq\label{dag}
\beta(T)\sp\dagger\pmatrix{1&0\cr 0&q}+\pmatrix{1&0\cr 0&q}
\beta(T) =0\, ,
\eeq
 and conversely, that for every $M\in M\sb 2\llp\Omega\sp1\! B\lrp$ satisfying 
(\ref{dag}) we can find unique
$\widetilde{a}$, $\widetilde{b}$, $\widetilde{t}$ such that $\beta (T) = M$
(cf.~Proposition~1 in~\cite{av}).
Thus we have proved the following:

\bpr\label{hermitian}
Let  
\mbox{$\cal C\sb B=\{\hb\!\in\!\mbox{\em Hom}\sb{\Bbb C}(U\sb q(2),\hO\sp1\! B)
\, |\;
\hb\ \mbox{satisfies (\ref{beta2}) for some}\;\widetilde{a},
\widetilde{b},\widetilde{t}\in\hO\sp1\! B\}$} and 
$\cal C\cal C(B\sp2)$ denote the space of Hermitian connections on $B\sp2$
(see Point~1 and Point~3 of Definition~\ref{pmdef}), where
\beq\label{beta2}
\beta(a\sp ka\sp{*l}b\sp m b\sp{*n}t\sp
pt\sp{*r}):= \left\{\begin{array}{ll}
0 &\mbox{for $m+n\geq 2$ }\\
\widetilde{b} & \mbox{for $m=1$, $n=0$}\\
\widetilde{b}\,\sp* &\mbox{for $m=0$, $n=1$
 }\\
(k-l)\widetilde{a}+(p-r)\widetilde{t}
 & \mbox{for $m=n=0\, ,$}
\end{array} \right .
\eeq
and $\langle\; ,\;\rangle$ is
given by the formula $\langle\xi , \zeta\rangle =\xi\sp\dagger 
\mbox{\scriptsize $\pmatrix{1&0\cr 0&q}$}\zeta$.
The map 
\[
\psi :\cal C\sb B\ni\hb\longmapsto d+\hb (T)\in\cal C\cal C(B\sp2)
\]
is a bijection.
\epr

\bco\label{hercor}
Let $P(B,U\sb q(2))$ be a trivial quantum principal bundle with the
universal differential calculus. There exists a one-to-one correspondence
between the elements of $\cal C\cal C(B\sp2)$, i.e.~the universal
calculus connections on $B\sp2$ that are compatible with the Hermitian
metric given by the matrix $\mbox{\scriptsize $\pmatrix{1&0\cr 0&q}$}$, 
and the strong connections
on $P(B,U\sb q(2))$ whose `pullback' on the base space $B$ (given
by (\ref{betadef})) satisfies (\ref{beta2}).
\eco

This way, we obtain the Yang--Mills theory of connections compatible with the
\mbox{$q$-d}ependent Hermitian structure on $B\sp 2$. Now, in order to handle
the critical points of $YM$ the same way they were dealt with in \cite{ri},
let us assume (as in \cite{ri}, p.535--6) that $B$ is a smooth dense 
$*$-subalgebra of a $C\sp*$-algebra, and that it is equipped with a faithful
invariant trace~\ctb . As was argued in Section~1 of
\cite{conri}, the Yang--Mills  functional
does not depend on the choice of
a Hermitian metric --- we can gauge $q$ out of the picture. 
Hence, the critical points of the \mbox{$U\sb q(2)$--Yang--Mills} action
functional are simply the critical points
of the Yang--Mills action (see p.536 in~\cite{ri}), if there are any,
`rotated' by the appropriate $q$-dependent gauge
transformation (see Section~1 in~\cite{conri}). 
(More explicitly, as the Hermitian metric is given by
$\mbox{\scriptsize $\pmatrix{1&0\cr 0&q}$}$, 
the corresponding gauge transformation is
$\nabla\to\mbox{\scriptsize $\pmatrix{1&0\cr 0&\sqrt{q}}$}\sp{-1}
\nabla\mbox{\scriptsize $\pmatrix{1&0\cr 0&\sqrt{q}}$}$.)
Consequently, we have:

\bco\label{moduli}
In the above described setting, the $U\sb q(2)$
and $U(2)$--Yang--Mills theories have
 the same moduli spaces of critical points (cf.\ Section~4 in
\cite{ri} and p.582 in~\cite{conbook}). 
\eco

Another way to remove $q$
from the picture is to alter (\ref{beta2}) by replacing $\widetilde{b}\,\sp*$ by
$q\widetilde{b}\,\sp*$. Then, however, we would lose the
geometrical interpretation of the
 action of the $\mbox{$q$-deformation}$ of $U(2)$
on the space of
compatible connections as the action of
 the gauge transformation. Also, the formula
$\beta(T\sp*) = \beta(T)\sp* $ 
would no longer be
true. (Caution: at least in the general case, we cannot
claim that $\beta$ is a \mbox{$*$-morphism} even when it commutes
with the $*$ on the generators.) In any case, we can see
 that the Yang--Mills theory remains unchanged
for any $q\in{\Bbb R}\sb +$. A~similar
 situation was discussed in the context of
quantum group gauge theories on classical
 spaces in the last two paragraphs of
Section~2 in \cite{bmc}.

\bre\label{sl}\em
The reason for employing in the considerations
above $U\sb q(2)$ rather than 
$SU\sb q(2)$ is that, when using $SU\sb q(2)$, formula
(\ref{beta2}) entails
$Tr\,\beta (T) = 0$ for all $\beta$, and although the tracelessness of
 $\beta (T)$ is automatically
preserved by the entire $GL(B\sp2)$ in the
classical case, we cannot claim the
same in general (cf.~Introduction and Section~3 in~\cite{arar}). 
Nor can we claim that the
tracelessness of $\beta (T)$ is preserved by the 
gauge group $U(B\sp2)$ of unitary automorphisms of $B\sp 2$
(see Remark~\ref{ua}). Besides,  $U\sb q(2)$-connections
given by (\ref{beta2}) can be regarded as $\langle \; ,\;\rangle$-compatible
 connections, and vice versa, which makes a clear analogy with
the classical situation, where Hermitian metrics are $U(2)$-structures
(cf.\ p.13 in \cite{tr}), and $U(2)$-connections are automatically compatible
 with the corresponding Hermitian structures (cf.~p.94--5 in~\cite{tr}).
\hfill{$\Diamond$}\ere

 Finally, let us mention that
an alternative approach would be to work with a non-universal differential
calculus instead of  assuming (\ref{betafor}), which is put in
the theory by hand. But that is yet another story.
\bigskip
\bigskip

{\parindent=0pt
{\bf\Large Appendix}\vspace*{-6pt}}
\appendix
\section{Quantum Associated Bundles}
\bde [cf.\ A.3 \cite{bmq}]\label{def.assoc.bundle}
 Let $(P,A,\Delta\sb R,N\sb P, M\sb A)$ be a
 quantum principal  bundle and
 $(V,\rho\sb R)$ be a right $A\sp{op}$-comodule algebra (see the remark below).
Also, let $\Delta\sb E$ be the
 homomorphism  from $P\ot V$ into $P\ot
V\ot A$ given by 
\[
\Delta\sb E = \llp id\ot id\ot (m\cc\tau )\lrp\ci (id\ot\tau\ot
id)\ci (\dr\ot\rho_{R})\, ,
\]
and $E$ be the space of all right-invariant
elements of $P\ot V$, i.e.~
\[
E=\{t\in P\ot V\; |\; \Delta\sb E
t=t\ot 1\}\, .
\]
Then $(E,P,V,\rho\sb R) $
is called the quantum fiber bundle associated
by $\rho\sb R$ to the quantum
principal bundle $P$.
\ede

\bre\label{bialgebra}\em
Note that the notion of a comodule algebra makes sense for any bialgebra.
In the preceding definition, we do not assume that $A\sp{op}$ is a Hopf algebra.
Such an assumption is equivalent to an assumption that the antipode of $A$ is 
bijective 
 --- a restriction that is unnecessary here. (Contrary to
\cite{bmq}, we use $S$ rather than $S\sp{-1}$.)
Observe also that the homomorphism
$\Delta\sb E$ makes $P\ot V$ a right 
\mbox{$A$-comodule}.
\hfill\mbox{$\Diamond$}\ere

We call $E$ a quantum fiber bundle instead of
a quantum vector bundle, as it is called in~\bmq .
The only difference, however, between Definition~A.3 \cite{bmq} and
 Definition~\ref{def.assoc.bundle} is an
extra twist in the formula defining $\Delta \sb E$
(i.e.~$m$ is replaced by $m\ci\tau$;
for comparison see also Corollary~4.14 in \cite{mp} and Section~5 in \cite{bk}).
An advantage of introducing such a
change is that, from the point of view of further constructions,
it is consistent with defining strongly horizontal differential 
forms as elements of
$\hO (B)P$ rather than $P\hO (B)$. As to using the word ``fiber" rather than
``vector",\footnote{I owe noticing this point to Marc Rieffel.}
the reason for doing so is that, since in order to
define section-valued differential forms $\Gamma\sp*(E)$
 (Definition~\ref{secdef}) or to prove the propositions
presented here it is unnecessary to assume
that $V$ is a quantum vector space
(quadratic algebra; see \cite{ma}), we formulate everything
without this assumption, and treat $V$ as a `quantum fiber'.
(See the paragraph below Corollary~\ref{D2cor} for how, having a linear
structure on a fiber, one can reproduce the familiar classical situation of
vector valued differential forms.)
The following are the reformulations of the corresponding
statements in \bmq . Except for the last parts of the proofs of
Proposition~\ref{asssecprop} and Proposition~\ref{70},
the remaining proofs in this section are
straightforward modifications of the corresponding proofs in \cite{bmq}.

\bpr  [{\rm cf.\ Lemma~A.4 \cite{bmq}}]\label{a4}
\hsp{-1mm}The spaces $E$ and $B\ot 1$ (see
Definition~\ref{qpbdef}) are
subalgebras of $P\ot V$ and $E$ respectively.
\epr
{\it Proof.} Analogous to the proof of Lemma~A.4\bmq .
\hfill{$\Box$}
\bigskip

Many statements below concern differential
forms of an arbitrary degree. Therefore, we need to assume that
we have a quantum principal bundle $P(B,A)$
equipped with a differential algebra $\hO (P)$ such that \op\ is
the first order differential calculus of $P(B,A)$.
\bde [{\rm cf.\ (69)\bmq}]\label{asssec}
A $k$-linear map $s : E\ra\Omega\sp{n}\!
 (\! B)$, with $n\in\{0\}\cup {\Bbb N}$,
 satisfying
$\forall\, b\!\in\! B:\, s\ci\widetilde{R}\sb{b\ot 1} = R\sb b\ci s$ and 
$s(1\te 1)=\delta\sb{n,0}\,$, where $\widetilde{R}$ and $R$
 denote the corresponding multiplications on the right,
is called a differential form with values in
 sections of the quantum fiber bundle $E$.
(For $n=0$, $s$ is simply called a section of a quantum
fiber bundle.)\label{secdef}
The space of all such $n$-forms will be denoted
 by $\Gamma\sp{n}(E)$.
\ede

\bpr  [cf.\ Proposition~A.5
\cite{bmq}]\label{asssecprop}
 Let $\phi\in ST\sb\rho\llp V,\Omega\sp n(P)\lrp$,
\nin , be such that
\mbox{$\phi(1)\! =\!\delta\sb{\deg \varphi ,0}$} (the space of
all such forms will be denoted by
$\overline{ST}\sb\rho\llp V,\Omega\sp n(P)\lrp$) and let\linebreak
\mbox{$s:=m\sb{\Omega(P)}\ci\tau\ci (id\te\phi)|\sb E\,$}.
Then, if $\deg \phi =0$ or if $P(B,A)$ is a trivial bundle,
\mbox{ $s\in\Gamma\sp{\deg\varphi}(E)$}.
\epr
{\it Proof.} To begin with, observe that for any $\phi\in
\overline{ST}\sb\rho\llp V,\Omega\sp n(P)\lrp$, \nin\ and $b\in B$,
 we have
\mbox{$s\cc\widetilde{R}\sb{b\ot 1} = R\sb b\cc s$}
 and $s\, (1\ot 1) =\hd\sb{\deg s,0}\,$,
so that it only remains to be shown that
 $s(E)\inc\hO\sp{\deg \varphi}(B)$.
Note also that, if $\sum\sb ip\sp i\te v\sp i\in E$, we have
\[
\llp\dsr\cc m\sb{\Omega (P)}\cc\tau\cc (id\te\phi )\lrp\,
(\mbox{$\sum\sb i$}p\sp i\te v\sp i)
=\lblp\llp m\sb{\Omega (P)}\cc\tau\cc (id\te\phi )\lrp\,
(\mbox{$\sum\sb i$}p\sp i\te
 v\sp i)\lbrp\ot 1\, .
\]
(We write the sums because we cannot claim that $E$ is
spanned by simple tensors.) 
Now, since it is clear that
$s(E)\inc \hO\sp{\deg \varphi}\sb{shor}(P)$, we can
conclude that $s$ takes values in strongly horizontal right-invariant
differential forms on $P$.
Hence, if $\deg \phi = 0$, then $s(E)\inc B$, and it follows
that $s\in\hG\sp 0(E)$. On the other hand,
if there exists a trivialization
of a quantum principal bundle $P(B,A)$,
by the same reasoning as was used to justify
formula (\ref{shorlem}),
we can infer that strongly horizontal right-invariant differential
forms must be elements of $\hO (B)$. Consequently,
\mbox{$s\in\hG\sp{\deg \varphi}(E)$}
as claimed. (Warning: As of now,
we do not know whether or not it is always true
that every right-invariant strongly
horizontal form on $P$ is a form on~$B$. Hence, as a precaution,
 we assume the triviality of
$P(B,A)$ when dealing with forms of degree bigger
than 0.)
\hfill{$\Box$}
\bigskip

For the remaining two propositions we will assume
 that $P(B,A)$ is a trivial quantum
principal bundle with a trivialization~$\Phi$.

\bpr  [cf.\ (71) and (72) in \cite{bmq}]\label{71}
Let $\Phi\sb E :=\llp (\Phi\cc S)\ot
id\lrp\ci\tau\ci\rho\sb R$.
Then
\[
{\cal I}:=(m\sb P\te id)\ci (id\te\Phi\sb E)
\]
 is a left $P$-module automorphism of $P\ot V$.
 Its inverse is given by
\[
\widetilde{\cal I}:=(m\sb P\te id)\ci\llp id\te (\Phi\sp{-1}\cc S)
\te id\lrp
\ci\llp id\ot(\tau\cc\rho\sb R)\lrp\, .
\]
 Moreover,
\mbox{$\Phi\sb E (V)\inc E$} and \mbox{$\widetilde{\cal I}(E)=B\ot V$}.
\epr
{\it Proof.}
{\parindent=0pt\it\smallskip

1.} It is straightforward to check that 
$\widetilde{\cal I}\cc{\cal I}=id={\cal I}\cc\widetilde{\cal I}$. 
Since ${\cal I}$ is, clearly,
a left $P$-module homomorphism, we can now conclude
that ${\cal I}$ is a left $P$-module automorphism of $P\ot V$. 
{\parindent=0pt\bigskip

{\it 2.}} A direct sigma notation calculation shows that
$\hD\sb E\ci\Phi\sb E=\Phi\sb E\ot 1$, whence $\Phi\sb E(V)\inc E$.
{\parindent=0pt\bigskip

{\it 3.}}
Finally, we must show that $\widetilde{\cal I}(E)= B\ot V$.
To do so, first we need:

\ble[cf.\ 3.1\cite{sch1} and (72)\bmq ]\label{Elem}
For every $\sum\sb ip\sp i\ot v\sp i\in E$,
\[
\sum\sb ip\sp i\0\ot v\sp i\ot p\sp i\1
=\sum\sb ip\sp i\ot v\sp i\0\ot S(v\sp i\1 )\, .
\]
\ele
{\it Proof.} This lemma can be verified by applying
\[
(id\te id\te m)\ci (id\te id\te S\te id)\ci (id\te\rho\sb R\te id)
\]
to both sides of the equality $\sum\sb i
p\sp i\0\te v\sp i\0\te v\sp i\1 p\sp i\1 =
\sum\sb ip\sp i\te v\sp i\te 1$.
\hfill{$\Box$}
\bigskip

Now, with the help of the above lemma, one can see that
$\widetilde{\cal I}|\sb E=\mbox{s}\sb\Phi\ot id$ (see (\ref{sphi})),
whence $\widetilde{\cal I}(E)\inc B\ot V$. Furthermore,
as $B\ot 1$ is a subalgebra
of $E$ (see Proposition~\ref{a4}), and $\Phi\sb E(V)\inc E$,
 we also have
${\cal I}(B\te V)\inc E$.
Thus, since $\widetilde{\cal I}\ci{\cal I}=id$, it
follows that\linebreak $\widetilde{\cal I}(E)=B\ot V$.
\hfill{$\rule{7pt}{7pt}$}
\bigskip

Closing our list of propositions is:
\bpr\label{70}
Let $\widetilde{\Phi}\sb
E:=(\Phi\sp{-1}\!\ot
id)\ci\tau\ci\rho\sb R$ (cf.\ (73)\cite{bmq}).
Then $\widetilde{\Phi}\sb E (V)\inc E$, and
\[
\Psi : \Gamma\sp{*}(E)\ni
 s\longmapsto (s\ci
\widetilde{\Phi}\sb E)*\sb\rho\Phi\in
 \overline{ST}\sb\rho
\llp V,\Omega\sp{*}(P)\lrp
\]
is a bijection.
\footnote{
This isomorphism corresponds to the formula for a pseudotensorial
\mbox{0-form} on $P$ that one could obtain by combining the very last formula
in the proof of Corollary~A.8 and formula (74) in \bmq .}
 Its inverse is given by
\[
\hsp{22mm}\widetilde{\Psi}:
\overline{ST}\sb \rho\llp V,\Omega\sp{*}(P)\lrp\ni
\phi\longmapsto m\sb{\Omega(P)}\ci\tau\ci(id\ot
 \phi)|\sb E\in\Gamma\sp{*}(E)\, .\hsp{10mm}
\mbox{ (cf.\ (70)\cite{bmq})}
\]
\epr
{\it Proof.}
{\parindent=0pt\it\smallskip

1.}
Using formula (28)\bmq\ (see the proof of Corollary~\ref{betainvcor}),
one can prove the inclusion
$\widetilde{\Phi}\sb E(V)\inc E$ in the same way as the inclusion
$\Phi\sb E(V)\inc E$, i.e.~by a direct sigma notation calculation.
(Observe that $\widetilde{\Phi}\sb E$ and
$\Phi\sb E$ coincide in the classical case.)
{\parindent=0pt\it\smallskip

2.} We already know from Proposition~\ref{asssecprop}
 that $\widetilde\Psi$ indeed takes values
in $\hG\sp*(E)$. Furthermore, it is clear that
\[
\mbox{\large $\forall\: $}s\in\hG\sp n(E),\,\nin\, :\; \Psi(s)(V)
\inc\hO\sp n\sb{shor}(P)\;\;\;\mbox{and}\;\;\; \Psi(s)(1)
=\hd\sb{n,0}\, .
\]
Proceeding much as in the first calculation in the proof of 
Proposition~A.7\bmq ,we can show that,
for an arbitrary $s\!\in\!\hG\sp n(E),\, \nin$, we have  
\[
\dsr\ci\Psi(s) =
\llp\Psi(s)\te id\lrp\ci\rho\sb R\, .
\]
Hence the inclusion 
\mbox{$\Psi\llp\hG\sp*(E)\lrp\inc\overline{ST}\sb\rho\llp V,\Omega\sp{*}(P)\lrp$}
follows.

{\parindent=0pt\it
3.}  For an arbitrary $s\in\hG\sp n(E)$, \nin\ and
$\sum\sb ip\sp i\te v\sp i
\in E$, we have
\bea\hsp{4mm}
&&
(\widetilde{\Psi}\cc\Psi )(s)(\mbox{$\sum\sb i$}p\sp i\te v\sp i)
\\ && =\sum\sb i
\Lblp m\sb{\Omega (P)}\ci\tau\ci\Llp id\ot\llp
(s\cc\widetilde{\Phi}\sb E)*\sb\rho\Phi\lrp\Lrp\Lbrp
(p\sp i\te v\sp i)
\\ && =\sum\sb i
(s\cc\widetilde{\Phi}\sb E)(v\sp i\0 )\,\Phi (v\sp i\1 )\, p\sp i
\\ && =\sum\sb i
(id*\sb{\cal R}\Phi\sp{-1})\llp
(s\cc\widetilde{\Phi}\sb E)(v\sp i\0 )\,\Phi (v\sp i\1 )\, p\sp i\lrp
\hsp{40.2mm}\mbox{(Im$\,(\widetilde{\Psi}\cc\Psi)(s)\inc\hO (B)$)}
\\ && =\sum\sb i
(s\cc\widetilde{\Phi}\sb E)(v\sp i\0 )\,
\mbox{s}\sb\Phi\llp\Phi (v\sp i\1 )p\sp i\lrp
\hsp{67.5mm}\mbox{(see (\ref{sphi}), cf.\ (\ref{idr}))}
\\ && =\sum\sb i
s\Lblp \widetilde{\Phi}\sb E(v\sp i\0 )\,\Llp\mbox{s}\sb\Phi\llp\Phi
 (v\sp i\1 )p\sp i\lrp\te 1\Lrp\Lbrp
\hsp{57mm}\mbox{($s\cc\widetilde{R}\sb{b\ot 1}=R\sb b\cc s$)}
\\ && =\sum\sb i
s\lblp\llp\Phi\sp{-1}(v\sp i\1)\te v\sp i\0\lrp
\llp\Phi (v\sp i\2)p\sp i\0\Phi\sp{-1}(v\sp i\3 p\sp i\1 )
\te 1\lrp\lbrp
\\ && =\sum\sb i
s\llp p\sp i\0\Phi\sp{-1}(v\sp i\1 p\sp i\1)\te v\sp i\0\lrp
\\ && =
(s\ci\mbox{\sf F})(\mbox{$\sum\sb i$}\, p\sp i\0\te v\sp i\te p\sp i\1 )
\\ && =
(s\ci\mbox{\sf F})\llp\mbox{$\sum\sb i$}\,
p\sp i\te v\sp i\0\te S(v\sp i\1 )\lrp
\hsp{68.5mm}\mbox{(see Lemma~\ref{Elem})}
\\ && =
s\Lblp\sum\sb i p\sp i\,\Phi\sp{-1}\llp
v\sp i\1 S(v\sp i\2 )\lrp\te v\sp i\0\Lrp\Lbrp
\\ && =
s(\mbox{$\sum\sb i$}p\sp i\ot v\sp i)\, ,
\hsp{100.8mm}\mbox{($\,\Phi\sp{-1}(1)=1\,$)}
\eea
where
\[
\mbox{\sf F}:=(m\sb P\te id)\ci
\llp id\te(\Phi\sp{-1}\!\cc m)\te id\lrp
\ci(id\te id\te\tau)\ci\llp id\te(\tau\cc\rho\sb R)\te id\lrp\, .
\]
Furthermore, it is straightforward to check that
$\Psi\ci\widetilde{\Psi}=id$.
Hence, $\widetilde{\Psi}$ is the inverse of $\Psi$, as needed.
\hfill{$\Box$}

\section{Axiomatic Definition of a Frame Bundle}

The theorem below allows an axiomatic definition of a frame bundle. 
This theorem should
have been proven many years ago. Nevertheless, since no appropriate
reference has been found, we include a proof.

\bth\label{afbthe}
Let $M$ be a smooth manifold with $\dim M = n \in{\Bbb N}$. 
A principal fiber bundle 
$P\mbox{\large$($}M\mbox{\large $,$}GL(n,{\Bbb R})\mbox{\large$)$}$ 
is isomorphic to the frame bundle
$FM$ if and only if there exists a smooth ${\Bbb R}\sp{n}$-valued 
$id$-equivariant \mbox{1-form} $\theta$ on $P$ such that $\ker \,
 \theta = \ker \,  \pi_{*}$ , where
$\pi : P\ra M$ is the bundle projection.
\ethe

\bre\label{footnote}\em
Here, by an isomorphism of two principal fiber bundles $P(M,G)$ and
$P'(M,G)$ we understand a diffeomorphism $f :
 P\ra P'$ satisfying $\pi '\ci f = \pi$ and 
$
f\ci R_{g}= R'_{g}\ci f
$
for any $g\in G$.~\hfill\mbox{$\Diamond$}
\ere

{\it Proof of Theorem~\ref{afbthe}.}
It is clear that if $P\llp M\mbox{\large$,$}GL(n,{\Bbb R})\lrp$ is
isomorphic to $FM$, then
the canonical $\mbox{1-form}$ on $FM$
yields, via the bundle
isomorphism, the desired \mbox{1-form} on $P$.
 Conversely, assume that
$P$ admits $\theta$ satisfying the above
conditions. Then, remembering that a coframe at any point
$m\!\in\! M$ is a linear
isomorphism from $T\sb mM$ to ${\Bbb R}\sp n$, we
can construct the following map:
\bea
&& J :
 P\stackrel{\check{J}}{\lra}F\sp{*}M\lra
FM\\
&& J : p\longmapsto\sigma\sp{*}\theta_{p}\longmapsto(
\sigma\sp{*}\theta_{p})\sp{-1},
\eea
where $\sigma$ is any smooth section of $P$
 verifying $(\sigma
\ci\pi )(p) = p$. Note that, due to the
assumption  $\kr\,\theta
= \kr\,\pi_{*}$ , the map $\check{J}$ indeed takes values
in the coframe bundle
$F\sp{*}M$  and does not depend on the choice
of $\sigma$. Thus, $J$ is uniquely and
well-defined. Let \mbox{$\widetilde{R} :
 FM\times GL(n,{\Bbb R})\ni (e,g)\mapsto e\cc g\in FM$} be the
right action and $\widetilde{\pi} :
 FM\ra M$ be the bundle projection. Evidently,
 $\widetilde{\pi}\ci J = \pi$ and,
for any \mbox{$g\!\in\! GL(n,{\Bbb R})$},
$\widetilde{R}_{g}\ci J = J\ci R_{g}$ ,
 whence $J$ is a bijection. Now, let $p$
be any point in $P$. There always exists
an open neighborhood $U$ of the point
$\pi (p)$ over which we can pick smooth
 sections $\chi : U\ra P$ and
$\widetilde{\chi} : U\ra FM$. Since
$p$ is arbitrary, to prove that $J$ is
smooth,
it suffices to show that the map
$T_{\tilde{\chi}}\ci J\ci
T_{\chi}\sp{-1} : U\times
GL(n,{\Bbb R})\ra U\times GL(n,{\Bbb R})$,
 where $T_{\tilde{\chi}}$ and
$T_{\chi}$ are local trivializations associated
 with $\widetilde{\chi}$ and $\chi$
respectively, is smooth. We have
\beq
(T_{\tilde{\chi}}\ci J\ci T_{\chi}\sp{-1})(u,g)
 =\lblp u\;
\mbox{\large\boldmath $,$}\;\llp\sigma\sp{*}\theta_{\chi (u)}\ci
\widetilde{\chi}(u)\lrp\sp{-1}\ci g\lbrp\, .
\label{f}\eeq
Since $\sigma\sp{*}\theta_{p}$ does not
 depend on the choice of $\sigma$, for any
$p\in \chi (U)$, we can pick $\sigma = \chi$.
 Taking (\ref{f}) into account, one can see that
 the smoothness of $J$ follows from the smoothness
of the map
\[
\ell : U\ni u\longmapsto\theta_{\chi
(u)}\ci\chi_{*}\ci\widetilde{\chi}(u)\in
 GL(n,{\Bbb R}).
\]
Furthermore, $J\sp{-1}$ is also smooth because 
\mbox{$(T_{\chi}\ci J\sp{-1}\ci
 T_{\tilde{\chi}}\sp{-1})(u,g) =
\mbox{\boldmath $($}u\,\mbox{\boldmath $,$}\,\sigma\sp{*}\theta_{\chi(u)}\ci
\widetilde{\chi}(u)\ci g\mbox{\boldmath $)$}$}.
Hence $J$ is a bundle isomorphism and
 $P\mbox{\large$($}M\mbox{\large $,$}GL(n,{\Bbb R})\mbox{\large$)$}$
 and $FM$ are isomorphic, as \linebreak
claimed.\hfill{$\Box$}\ \\

An advantage of such an axiomatic definition of
 a frame bundle is that it is,
at least a priori, translatable into
the language of quantum principal
 bundles. More precisely, we could
 try to define a quantum frame bundle
as a quantum  bundle $P(B,A)$ equipped
 with a horizontal (or strongly horizontal)
form $\theta : V\hookrightarrow\op$
 satisfying $\dr\ci\theta =
(\theta\ot id)\ci\rho_{R}$, where
$(V,\rho_{R})$ is an appropriate right
$A\sp{op}$ (or $A$) comodule algebra.
(An example of such a construction for the case of a classical group
has been provided in~\cite{md5}; in particular, see Section~4 of~\cite{md5}.)
It seems interesting to consider in this context bundles over the  
quantum sphere $S\sp{2}_{q}$ (see \cite{po} or~\cite{bmq}) with 
$A = GL_{q}(2,\Bbb R)$ (see p.161 in~\cite{ta}), 
$V$~a~quantum plane (see~\cite{ma}) and $\rho_{R}$ the standard
right coaction (see Appendix~B in \cite{bmq}).\\

\footnotesize
{\it Acknowledgments.}
This work 
was in part supported by the NSF grant 1--443964--21858--2.
Writing up the revised version was partially supported by the KBN grant
\mbox{2 P301 020 07} and by a visiting fellowship at 
the International Centre for Theoretical Physics in Trieste.
I am particularly indebted to Tomasz
 Brzezi\'{n}ski and Marc Rieffel, whose
manifold and generous assistance facilitated
the preparation of this article. I also would like to
express my gratitude to Shahn Majid, Markus Pflaum, Nicolai Reshetikhin, 
Philip Ryan, Shuzhou Wang
and Stanis\l aw Zakrzewski for helpful discussions, and
to Shoshichi Kobayashi and Joseph Wolf for consultations
on Theorem~\ref{afbthe}.


\begin{thebibliography}{99}
\parindent0pt

\vspace*{-0mm}\bibitem{abe} Abe, E.:~ Hopf Algebras.   
 Cambridge University Press  
1980

\vspace*{-0mm}\bibitem{arar} Aref'eva, I.Ya., Arutyunov, G.E.:~  
Uniqueness of $U\sb q(N)$ as a quantum gauge group and representations of its
differential algebra.  
\mbox{\sf (hep-th/9305176)}

\vspace*{-0mm}\bibitem{av} Aref'eva, I.Ya., Volovich, I.V.:~  
Quantum group gauge fields.  
Mod.\ Phys.\ Lett.\ A~ {\bf 6} (10), 893--907 (1991)

\vspace*{-0mm}\bibitem{booss} Booss, B., Bleecker, D.D.:~  
Topology and Analysis. The Atiyah--Singer Formula
and \mbox{Gauge--Theoretic} Physics.  
Springer--Verlag 
1985

\vspace*{-0mm}\bibitem{aki} Bourbaki, N.:~ 
Alg\`{e}bre Homologique.
 Masson 
1980

\vspace*{-0mm}\bibitem{tbp} Brzezi\'{n}ski, T.:~ 
Differential Geometry of Quantum Groups and Quantum Fibre Bundles. 
Ph.D.\ thesis, Cambridge University 1994

\vspace*{-0mm}\bibitem{tbt} Brzezi\'{n}ski, T.:~ 
Translation Map in Quantum Principal Bundles.  
to appear in J.\ Geom.\ Phys.
\mbox{\sf (hep-th/9407145)} 

\vspace*{-0mm}\bibitem{bmc} Brzezi\'{n}ski, T., Majid, S.:~	 
Quantum Group Gauge Theory on Classical Spaces.   
Phys.\ Lett.~B~ {\bf 298},  339--43 (1993)
\mbox{\sf (hep-th/9210024)}

\vspace*{-0mm}\bibitem{bmq} Brzezi\'{n}ski, T., Majid, S.:~ 
``Quantum Group Gauge Theory on Quantum Spaces"~  
Commun.\ Math.\ Phys. {\bf 157}, 591--638 (1993)
\mbox{\sf (hep-th/9208007)}; 
erratum Commun.\ Math.\ Phys. {\bf 167}, 235 (1995)

\vspace*{-0mm}\bibitem{bk} Budzy\'{n}ski, R.J., Kondracki, W.:~   
Quantum Principal Fiber Bundles: Topological Aspects.   
\mbox{\sf (hep-th/9401019)} 

\vspace*{-0mm}\bibitem{conact} Connes, A.:~ 
The Action Functional in Non-Commutative Geometry.  
Commun.\ Math.\ Phys. {\bf 117}, 673--83 (1988)

\vspace*{-0mm}\bibitem{conbook}  Connes, A.:~
Non-Commutative Geometry. Academic Press 1994

\vspace*{-0mm}\bibitem{conri} Connes, A., Rieffel, M.:~ 
Yang--Mills for Non-Commutative Two-Tori.  
Contemp.\ Math. {\bf 62}, 237--66 (1987)

\vspace*{-0mm}\bibitem{coq} Coquereaux, R.:~ 
Noncommutative Geometry and Theoretical Physics.  
J.\ Geom.\ Phys. {\bf 6}, (3) 425--90 (1989)

\vspace*{-0mm}\bibitem{mgtqg} Doebner, H.D., Hennig, J.D., L\"{u}cke, W.~   
Mathematical Guide to Quantum Groups. In: Doebner, H.D., Hennig, J.D. (eds.)~ 
Quantum Groups.
Proceedings of the 8th International Workshop on Mathematical Physics, Clausthal,
West Germany, 19--26 July 1989. pp.~29--63. Springer--Verlag  
1990 
 
\vspace*{-0mm}\bibitem{dv} Dubois-Violette, M.:~ 
D\'{e}rivations et calcul diff\'{e}rentiel non commutatif.  
C.\ R.\ Acad.\ Sci.\ Paris {\bf 307} (S\'{e}rie I), 403--8 (1988)

\vspace*{-0mm}\bibitem{dvkmm} Dubois-Violette, M., Kerner, R., Madore, J.:~ 
Noncommutative Differential Geometry of Matrix Algebras.  
J.\ Math.\ Phys. {\bf 31} (2), 316--22 (1990)

\vspace*{-0mm}\bibitem{dkm} Dubois-Violette, M., Kerner, R., Madore, J.:~
Noncommutative Differential Geometry and New Models of Gauge Theory.  
J.\ Math.\ Phys. {\bf 31} (2), 323--30 (1990)

\vspace*{-0mm}\bibitem{md2} Durdevic, M.:~ 
Geometry  of Quantum  Principal Bundles  I.  
to appear in Commun.\ Math.\ Phys.  
\mbox{\sf (q-alg/9507019)}
   
\vspace*{-0mm}\bibitem{md3} Durdevic, M.:~  
Geometry of Quantum Principal Bundles II (Extended Version).    
 \mbox{\sf (q-alg/9412005)}

\vspace*{-0mm}\bibitem{md4} Durdevic, M.:~ 
Quantum Principal Bundles and Corresponding Gauge Theories.  
\mbox{\sf (q-alg/9507021)}

\vspace*{-0mm}\bibitem{md1} Durdevic, M.~ 
Quantum Principal Bundles. 
In: Proceedings of the XXII International
 Conference on Differential Geometric Methods
in Theoretical Physics. Ixtapa, Mexico 1993~ 
\mbox{\sf (hep-th/9311029)}

\vspace*{-0mm}\bibitem{md5} Durdevic, M.:~
On Framed Quantum Principal Bundles. 
\mbox{\sf (q-alg/9507020)}

\vspace*{-0mm}\bibitem{frt} Fadeev, L.D.,Reshetikhin,  N.Yu., Takhtadzhyan,  L.A.:~
Quantization of Lie Groups and Lie Algebras. 
Leningrad Math.\ J. {\bf 1} (1), 193--225 (1990)

\vspace*{-0mm}\bibitem{ha} Hazewinkel, M.:~
Formal Groups and Applications.  
Academic Press 1978

\vspace*{-0mm}\bibitem{sm} Majid, S.:~ 
Introduction to Braided Geometry and $q$-Minkowski space.
In:  Castellani, L., Wess, J. (eds.)~ 
Proceedings of the International School of Physics `Enrico Fermi'
CXXVII. IOS Press 1996~ 
\mbox{\sf  (hep-th/9410241)} 

\vspace*{-0mm}\bibitem{ma} Manin, Yu.I.:~ 
Quantum Groups and Noncommutative Geometry. 
 Publications du CRM 1561, Univ. de Montreal 1988

\vspace*{-0mm}\bibitem{mp} Pflaum, M.J.:~
Quantum Groups on Fibre Bundles. 
Commun.\ Math.\ Phys. {\bf 166} (2), 279--315 (1994)~ 
\mbox{\sf (hep-th/9401085)}

\vspace*{-0mm}\bibitem{po} Podle\'{s}, P.:~  
Quantum Spheres.
Lett.\ Math.\ Phys. {\bf 14}, 521--31 (1987)

\vspace*{-0mm}\bibitem{ri} Rieffel, M.:~
Critical Points of Yang--Mills for Noncommutative Two-Tori. 
J.\ Differential Geom. {\bf 31}, 535--46 (1990)

\vspace*{-0mm}\bibitem{sch1} Schneider, H.J.:~
Principal Homogenous Spaces for Arbitrary Hopf Algebras. 
Isr.\ J.\ Math. {\bf 72} (1--2), 167--95 (1990)

\vspace*{-0mm}\bibitem{sch2} Schneider, H.J.:~ 
Hopf Galois Extensions, Crossed Products, and Clifford Theory.
In:  Bergen, J., Montgomery, S. (eds.)~ 
Advances in Hopf Algebras.
 Lecture Notes in Pure and Applied Mathematics. 
{\bf 158}, Marcel Dekker, Inc. 1994

\vspace*{-0mm}\bibitem{shst} Shnider, S., Sternberg, S.:~
Quantum Groups. International Press Inc. 1993

\vspace*{-0mm}\bibitem{ms} Spera, M.:~
A Symplectic Approach to Yang--Mills Theory for Non Commutative Tori. 
Can.\ J.\ Math. {\bf 44} (2), 368--87 (1992)

\vspace*{-0mm}\bibitem{swe} Sweedler, M.E.:~ 
Hopf Algebras. Benjamin  
1969

\vspace*{-0mm}\bibitem{ta} Takhtadzhyan, L.A.
Lectures on Quantum Groups.
In: Introduction to Quantum Groups and Integrable Massive Models
 of Quantum Field Theory. pp.~69--197. 
 World Scientific 
1990

\vspace*{-0mm}\bibitem{tr} Trautman, A.:~ 
Differential Geometry for Physicists. 
Bibliopolis  
1984

\vspace*{-0mm}\bibitem{cmp} Woronowicz, S.L.:~
Compact Matrix Pseudogroups. 
Commun.\ Math.\ Phys.  {\bf 111}, 613--65 (1987)

\vspace*{-0mm}\bibitem{wor} Woronowicz, S.L.:~
Differential Calculus on Compact Quantum Pseudogroups (Quantum Groups). 
Commun.\ Math.\ Phys. {\bf 122}, 125--70 (1989)

\vspace*{-0mm}\bibitem{wz} Wu, K., Zhang, R.-J.:~ 
Algebraic approach to gauge theory and its noncommutative extension.  
Commun.\ Theor.\ Phys. {\bf 17} (2), 175--82 (1992)

\end{thebibliography}
\end{document}